\documentclass[sigplan,10pt]{acmart}
\renewcommand\footnotetextcopyrightpermission[1]{}
\usepackage{tikz}
\usepackage{amsmath}

\usepackage{tabularx}
\usepackage{makecell}
\usepackage{comment}
\usepackage{multirow}
\usepackage{array}
\usepackage{subcaption}
\usepackage{xspace}
\usepackage{xurl}
\usepackage{hhline}

\usepackage{tcolorbox}
\tcbuselibrary{skins, breakable}

\usepackage{todonotes}

\newcommand{\PLAN}[1]{
}

\newcommand{\takeawaybox}[1]{
\noindent \textbf{Takeaway:} {#1}
}

\newcolumntype{C}[1]{>{\hsize=#1\hsize\centering\arraybackslash}X}

\newcommand{\ABench}{AgentSysBench\xspace}

\usepackage{enumitem}
\setlist[itemize]{leftmargin=*, labelsep=4mm}

\usepackage{listings}
\lstdefinestyle{techstyle}{
  basicstyle=\ttfamily,
  frame=none,
  showspaces=false,
  showstringspaces=false
}
\newcommand{\techterm}[1]{\lstinline[style=techstyle]{#1}}

\begin{document}
\pagestyle{plain}

\title{From LLM Inference to Agentic Workloads: Characterization and Implications for Serving Systems}

\author{
{\rm Chaokun Chang}$^{1,*}$, {\rm Yukun Zhou}$^{1,*}$, {\rm Kaihua Fu}$^1$, {\rm Dakai An}$^1$, {\rm Tianyu Feng}$^1$, {\rm Hanfeng Lu}$^1$, {\rm Sheng Yao}$^1$, \\
{\rm Pu Guo}$^1$, {\rm Yinghao Yu}$^2$, {\rm Yizhou Shan}$^3$, {\rm Bo Li}$^1$, {\rm Binhang Yuan}$^1$, {\rm Wei Wang}$^1$ \\[4pt]
$^1${\rm Hong Kong University of Science and Technology} \\
$^2${\rm Alibaba Group}, $^3${\rm Bytedance}\\[2pt]
{\small $^*$Equal contribution.}
}

\begin{abstract}
Agentic applications are shifting AI serving from isolated model inference to long-running workloads in which LLMs coordinate tools, environments, and persistent state.
However, the system behavior of these workloads---where latency, cost, and bottlenecks arise---remains poorly characterized, leaving serving systems to rely on assumptions built for conventional inference.
We present \ABench, a benchmark suite and measurement toolkit with ten representative agentic applications and unified systems-level instrumentation.
Across controlled deployments and production traces, we identify six properties that distinguish agentic workloads from conventional LLM serving:
(1)~execution is heavyweight and stateful, with non-LLM components dominating latency in 5 of 10 applications and sandbox working-set memory peaking at 28 GB per session;
(2)~applications compose components with heterogeneous resource affinity---GPU-bound inference, memory-bound retrieval, CPU-bound sandboxes---whose task latencies diverge by up to 32$\times$;
(3)~bottlenecks shift across requests, models, and deployments;
(4)~production sessions hold state idle for minutes to hours between active steps;
(5)~a control-plane tax---auxiliary LLM calls and context overhead from tool schemas and observations---crowds out productive compute and context; and
(6)~production traces from three applications reveal heavy cross-request redundancy in search queries and web fetches, exposing a large caching opportunity.
Four design explorations demonstrate that these findings are actionable:
task-aware serving reduces latency by 29--40\%, communication-aware placement by up to 4.5$\times$,
state offloading reduces memory usage by 4.6$\times$, and tool-result caching removes 35.2\% of redundant search calls and saves 19.3\% of aggregate search latency.
\end{abstract}

\maketitle

\section{Introduction}

AI serving is shifting from isolated LLM inference to long-lived, tool-using
sessions that act on a user's behalf. Modern agentic applications use large language
models (LLMs) not just to generate text but to invoke tools, inspect external
environments, update persistent state, and iterate over intermediate
results~\cite{yao2023react,qin2023toolllm}. A single user request may trigger
retrieval, code execution, browser or GUI interaction, API calls, and dozens of
model invocations before producing a final answer, as in deployed coding,
research, and assistant
agents~\cite{claudecode,openai-codex,langchain-open_deep_research,shen2023hugginggpt}.
We refer to the resulting workloads---distributions of such session-level
executions---as \emph{agentic workloads}: they combine model inference, tool
execution, environment interaction, and state management under LLM-driven
control (\S\ref{sec:agentic-bkg}).

Yet the systems behavior of these workloads remains poorly characterized: no study instruments a
broad set of agentic applications under a unified serving stack and complements
controlled findings with production-trace behavior (\S\ref{sec:characterization-bkg}).
System developers thus cannot answer basic questions---what fraction of end-to-end latency
falls in tools versus models, how session state grows, whether bottlenecks shift across
requests and deployments, and how much work is productive versus control-plane overhead.
As a result, serving systems for agentic workloads are often designed using assumptions
inherited from conventional LLM inference~\cite{kwon2023efficient,zheng2024sglang},
even though agents exercise a much broader and more heterogeneous execution stack.

Prior work leaves three gaps. \emph{First}, recent agentic serving systems study
only a few applications, often one to three, under incompatible tool, environment,
model, and orchestration assumptions~\cite{ayo,ALTO,parrot,autellix}. These studies provide
valuable point solutions, but their workloads are difficult to compare across
systems and do not reveal which properties are fundamental versus
application-specific. \emph{Second}, capability
benchmarks such as SWE-bench~\cite{jimenez2023swe},
AgentBench~\cite{liu2023agentbench}, WebArena~\cite{zhou2023webarena}, and
OSWorld~\cite{osworld} measure task completion but record little systems-level
information: latency breakdowns, resource profiles, state footprints, tool-call
behavior, or cross-request redundancy. \emph{Third}, recent measurement studies of agentic
execution~\cite{yuan2025agentic,raj2025cpucentric,kim2025costdynamic,agentrace2025}
report useful observations on traces, CPU-side overhead, and reasoning cost, but
each runs on a single fixed serving stack and does not treat hardware
allocation, component placement, or production conditions as variables.

\begin{table}[!t]
\caption{Compact overview of the \ABench suite.}
\label{tab:suite-brief}
\centering
\footnotesize
\setlength{\tabcolsep}{3pt}
\begin{tabular}{p{0.19\linewidth}|p{0.27\linewidth}||p{0.19\linewidth}|p{0.27\linewidth}}
\hline
\textbf{Domain} & \textbf{Application} & \textbf{Domain} & \textbf{Application} \\ \hline
QA & RAG~\cite{RAG-Survey} & AI search & DeepResearch~\cite{langchain-open_deep_research} \\ \hline
Multimodal processing & HuggingGPT~\cite{shen2023hugginggpt} & Software\newline engineering & Mini-SWE~\cite{mini-sweagent} \\ \hline
Terminal use & Codex~\cite{openai-codex} & Browser use & WebAgent~\cite{webarena-verified} \\ \hline
Computer use & GUIAgent~\cite{osworld} & Tool-rich assistants & Claude Code~\cite{claudecode} \\ \hline
Office work & Openclaw~\cite{openclaw} & AutoResearch & Pi-AutoR~\cite{pi-autoresearch} \\ \hline
\end{tabular}
\vspace{-3mm}
\end{table}

To fill these gaps, we present \ABench, a benchmark suite and measurement
toolkit with three features. \emph{First}, it assembles ten representative agentic applications
(Table~\ref{tab:suite-brief}) spanning the execution patterns that shape systems
behavior---\emph{predefined} versus \emph{LLM-driven} control flow, diverse
tool/environment interaction, and \emph{long-running stateful
execution} (\S\ref{sec:benchmark-requirements}). \emph{Second}, it provides a \emph{unified harness}
that instruments LLM calls, tool invocations, and state operations, recording
latency, resource usage, data movement, live state footprint, token behavior,
and cost. \emph{Third}, it offers a \emph{modular serving stack} that deploys LLMs, embedding
models, vector databases, sandboxes, browser/GUI environments, search services,
and tool servers separately or co-located, enabling \emph{controlled study} of
provisioning, placement, and state management (\S\ref{sec:benchmark-suite}).

Using \ABench, we characterize workloads by combining \emph{controlled
experiments}---which isolate application, model, request, and deployment under unified instrumentation---with \emph{production traces} that reveal
phenomena synthetic workloads miss.
Our study covers 4,641 benchmark requests, 64,924 LLM calls, 118,274 tool calls in controlled experiments, and 178,799
production sessions in a single day.
Across both views, one theme recurs: model inference is no longer the sole
cost center. A large, often dominant share of latency,
memory, and cost arises in tools, environments, and long-lived session state, and
the dominant cost shifts across requests, models, and deployments. Efficient
agentic serving therefore requires \emph{coordinated management} of models, tools,
state, and communication, not model-centric optimization alone. We substantiate
this theme through six key findings, among which the first three are from the controlled experiments:

\vspace{2pt}
\noindent
(1)~{\bf Heavyweight execution with non-LLM dominance.} Agent execution is
\emph{long-running} and \emph{stateful}, issuing many LLM, tool, and environment
calls while accumulating \emph{live session state}---prompts, tool outputs, KV
cache, and artifacts---for the whole request. In 5 of 10 applications, tools and
environments dominate or co-dominate latency, so model-only optimization leaves
much execution time untouched. \emph{Implication:} the
serving stack must accelerate tool-heavy stages and manage session state, not
just model inference (\S\ref{sec:heavy}).

\vspace{2pt}
\noindent
(2)~{\bf Cross-stack heterogeneity.} A single application composes components with
\emph{divergent resource profiles}: GPU-bound LLMs, memory-bound vector databases,
CPU-bound sandboxes, and network-bound services. Even tasks sharing one component
differ by up to 32$\times$ in latency, causing severe head-of-line blocking under a single queue.
\emph{Implication:} provisioning and scheduling
must be heterogeneity-aware, matching each task's resource profile rather than
assuming uniform cost (\S\ref{sec:hetero}).

\vspace{2pt}
\noindent
(3)~{\bf Shifting bottlenecks.} The dominant component \emph{shifts} across
requests, models, tools, and deployments. Even a simple RAG pipeline shifts its
bottleneck between embedding and vector-database operations as inputs change.
\emph{Implication:} application-level averages mask this variance; serving needs
online, per-request adaptation, not static profiling (\S\ref{sec:dynamic}).

\vspace{4pt}
\noindent
Our production traces add three further findings that are difficult to uncover
with controlled benchmarks alone.

\vspace{2pt}
\noindent
(4)~{\bf Long idle-but-live intervals.} Production sessions often wait minutes to
hours between steps while holding conversational, tool, and environment
state---neither finished nor consuming compute. Treating them as live wastes
memory and disk; treating them as finished loses resumable state.
\emph{Implication:} serving systems must distinguish ``done'' from ``waiting''
to reclaim resources while preserving resumability, which we exploit for state
offloading (\S\ref{sec:case-study:pro-scheduling}).

\vspace{2pt}
\noindent
(5)~{\bf LLM control-plane tax.} Beyond productive work that deciding the next action
and producing outputs, tokens are spent on tool schemas, raw
observations, and safety checks. This \emph{control-plane tax} grows over multi-step
sessions and fills context, raising latency and cost.
\emph{Implication:} serving systems need context-budget-aware tool interfaces
and observation compression to recover capacity for useful work.

\vspace{2pt}
\noindent
(6)~{\bf Exploitable cross-request redundancy.} Among 373,678 search queries from a
production agentic-search application, 27\% of unique search queries recur and
account for 67.3\% of all search API calls.
In another production Openclaw-like application issuing 4,389 web fetches,
24\% of distinct fetched URLs recur and account for 64\% of fetch invocations.
This redundancy is invisible in single-task benchmarks.
\emph{Implication:} query- and object-level caching can
eliminate most redundant external calls, cutting latency and cost
(\S\ref{sec:case-study:caching}).

\vspace{4pt}
\noindent
These findings lead to \emph{actionable optimizations}. To show this, we conduct
four design explorations that yield significant performance gains:
task-disaggregated serving (29--40\% lower latency), agent-aware co-location (up
to 4.5$\times$), state offloading (4.6$\times$ less memory), and
tool-result caching (35.2\% fewer redundant search calls and 19.3\% less
aggregate search latency). These gains come from
simple, characterization-guided mechanisms rather than a full system optimization---evidence
that current workload-oblivious serving leaves substantial efficiency unclaimed.

We will release \ABench as open source so the community can build the next
generation of agent-serving systems with a workload-informed design.

\section{Background and Motivation}

Conventional LLM serving treats a token-generation request as the unit of work,
but an agentic application turns one user request into a long-running execution
that interleaves LLM calls, tool invocations, environment interactions, and
state updates. Serving such applications therefore requires understanding the
full execution induced by a request. This section
defines agentic workloads (\S\ref{sec:agentic-bkg}), explains why
characterizing them requires jointly controlling the workload and the serving
system (\S\ref{sec:characterization-bkg}), and derives the design requirements
for our benchmark suite (\S\ref{sec:benchmark-requirements}).

\subsection{Agentic Applications and Workloads}\label{sec:agentic-bkg}

\begin{figure}[t]
    \centering

    \includegraphics[width=1.0\linewidth]{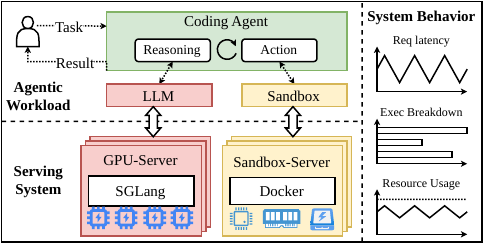}

    \caption{Agentic workload, serving system, and system behavior illustrated with a coding agent.}
    \label{fig:background}
    \vspace{-4mm}
\end{figure}

We use \emph{agentic application} to denote the user-visible AI application,
and \emph{agentic execution} to denote the runtime process induced by one user
request or \emph{session}. Unlike a chatbot request, which the serving system can
treat as a single stateless inference call, an agentic execution interleaves
LLM calls, tool invocations, environment interactions, and state operations
under runtime control decisions. Its structure forms an execution graph with
chains, branches, loops, and parallel sub-tasks; each request realizes one path
through this graph, and the events recorded along that path form an execution
trace. Figure~\ref{fig:background} (left) shows a coding agent as a running
example: an orchestrator drives a Reasoning--Action (ReAct) loop, repeatedly
dispatching to an LLM for reasoning and to a sandbox for action until a result
is produced.

For a systems study, the relevant object is not a single execution but a
distribution of executions, which we call an \emph{agentic workload} and
summarize by four factors, $W{=}\langle R,T,M,O\rangle$. The request distribution $R$
captures application scenario, task difficulty, and payload size. The tool and
environment set $T$ (e.g., sandboxes, browsers, and vector databases)
determines I/O behavior, state footprint, failure modes, and non-LLM
bottlenecks. The model choices and inference policies $M$ determine latency,
cost, token volume, context length, and cache behavior. The orchestration
structure $O$ (e.g., a predefined pipeline, a ReAct loop, or a planner--executor
design) shapes dynamicity, parallelism, and state retention. Together, these
factors determine execution length, resource demand, and cost;
\S\ref{sec:benchmark-suite} uses them to select representative
applications.

\subsection{The Gaps}
\label{sec:characterization-bkg}

\noindent{\bf Need for a systems benchmark suite.}
Serving systems for agentic applications are still designed largely with
assumptions inherited from conventional LLM
inference~\cite{kwon2023efficient,zheng2024sglang}. A primary reason is that the systems
community lacks a characterization of the workloads these applications induce.
Obtaining such a characterization is harder than profiling model inference:
agentic behavior cannot be characterized from the workload alone, because the
observed outcome also depends on how the workload is served. We model the
serving system as $S{=}\langle H,C,A\rangle$, where $H$ denotes hardware resources, including GPUs,
CPUs, DRAM, storage, and network bandwidth; $C$ denotes component-serving
mechanisms, i.e., how each component---the LLM engine, embedding or reranking
service, vector database, sandbox manager, browser or GUI environment, and
coordinator---is served; and $A$ denotes deployment architecture, i.e.,
whether components are co-located, separately containerized, or served by
remote cloud services. Figure~\ref{fig:background} (bottom) shows a concrete
instance: an SGLang LLM engine ($C$) on a GPU server ($H$) and a Docker
sandbox ($C$) on a physically separate server ($H$), communicating over the
network ($A$). A measured outcome is therefore an interaction $Y=\Phi(W,S)$
(Figure~\ref{fig:background}, right): the same application can appear
LLM-dominated with a slow model, tool-dominated with a slow sandbox,
state-dominated with a large live context, or network-dominated with remote
placement. If the serving stack is underspecified, a study may attribute to
the workload a bottleneck actually caused by hardware allocation, serving
policy, or placement; if the workload is underspecified, an optimization may
be tuned to one configuration with no evidence that it generalizes.

\vspace{2pt}
\noindent{\bf Prior work falls short.}
Existing work controls only one side of this interaction. Capability benchmarks
such as SWE-Bench~\cite{jimenez2023swe}, WebArena~\cite{zhou2023webarena},
OSWorld~\cite{osworld}, ToolBench~\cite{qin2023toolllm}, and
AgentBench~\cite{liu2023agentbench} provide realistic tasks and environments but
measure task completion on an implicit, uncontrolled serving stack, recording
little systems-level information---latency breakdowns, resource usage, state
footprints, or token accounting---so they cannot explain where serving cost
arises. Agent-serving systems~\cite{ayo,tan2024teola,ALTO,parrot,autellix} take
the opposite cut: each varies serving mechanisms but evaluates them on one to
three workflows under incompatible assumptions about requests, tools, models,
and orchestration, so their observations cannot be compared across studies or
shown to generalize beyond the evaluated applications. Recent measurement
studies~\cite{yuan2025agentic,raj2025cpucentric,kim2025costdynamic,agentrace2025}
come closest, instrumenting ReAct-style traces, tool costs, CPU-side
orchestration overheads, and dynamic reasoning costs, but each runs on a single
fixed serving stack rather than treating the stack as a \emph{variable}; a reported
bottleneck may therefore be specific to that stack's model engine, tool
implementation, hardware allocation, or placement policy.

\vspace{2pt}
\noindent{\bf Controlled study is insufficient.}
Even a jointly controlled study is incomplete, because some workload behavior
emerges only in production. Controlled experiments impose synthetic arrival
processes and run tasks to completion in isolation, so they cannot surface
realistic request arrivals and user think time, long-lived sessions that hold
state while idle, or redundancy across requests from different users. Our
production traces show that these phenomena are consequential: sessions wait
minutes to hours between steps while holding live state, and a majority of
external search calls repeat queries issued by earlier requests
(\S\ref{sec:production}). A characterization built only from controlled
benchmarks would miss the behaviors that drive memory occupancy and
external-call cost at scale.

\subsection{Design Requirements for the Benchmark Suite}
\label{sec:benchmark-requirements}

We believe a rigorous characterization must satisfy four needs: select
applications through an explicit workload model rather than an ad hoc
collection; jointly specify and control the workload $W$ and the serving
system $S$; instrument complete executions at the granularity of individual
LLM calls, tool invocations, and state operations; and complement controlled
findings with production traces. These needs translate into four requirements
for the benchmark suite.

\vspace{2pt}
\noindent\textbf{R1: Workload representativeness.}
The suite must cover diverse combinations of $R$, $T$, $M$, and $O$ rather
than a narrow selection of prompts or task labels. Narrow coverage confines
observations to application-specific phenomena; coverage across request
distributions, tool environments, model choices, and orchestration structures
is necessary to identify which properties are fundamental to agentic serving.

\vspace{2pt}
\noindent\textbf{R2: Controllable serving-system factors.}
The suite must let experiments vary hardware allocation ($H$),
component-serving mechanisms ($C$), and deployment architecture ($A$) under
controlled conditions. Without this capability, an observed bottleneck cannot
be attributed to the workload or the serving infrastructure, and an
optimization cannot be evaluated across deployment configurations.

\vspace{2pt}
\noindent\textbf{R3: Unified component-level instrumentation.}
Measurements across applications and deployments must be directly comparable,
and instrumentation must reach individual LLM calls, tool invocations, and
state operations---not only end-to-end outcomes---since attributing cost
requires per-component visibility into latency, resource usage, data movement,
token behavior, and state footprint.

\vspace{2pt}
\noindent\textbf{R4: Production complementarity.}
The suite must pair controlled benchmarks with traces collected from
production deployments, exposing the request arrivals, session lifetimes, and
cross-request redundancy that controlled experiments cannot reproduce. The
production study complements rather than directly validates the controlled
experiments by revealing additional deployment-only behavior.

\vspace{2pt}
\ABench is designed to meet all four requirements:
\S\ref{sec:benchmark-suite} describes the application collection (R1),
modular serving stack (R2), and measurement toolkit (R3), and
\S\ref{sec:production} presents the production-trace study that
addresses R4.

\section{Benchmarking Suite}\label{sec:benchmark-suite}

This section describes how \ABench turns the workload and serving-system models
from \S\ref{sec:agentic-bkg} and \S\ref{sec:characterization-bkg} into a
runnable benchmark suite. \ABench maps workload factors $W{=}\langle R,T,M,O\rangle$ to a
representative application collection (R1), exposes serving-system factors
$S{=}\langle H,C,A\rangle$ through a modular serving stack (R2), and records comparable traces
and resource time series across applications (R3). We then describe how new
workloads can be added and list the default settings used in our
characterization study.

\begin{table*}[t]
\caption{Systems-oriented overview of the benchmarking suite in \ABench.}
\label{tab:suite-app-ds}
\centering
{
\footnotesize
\setlength{\tabcolsep}{2.5pt}
\begin{tabular}{l|l|l|l|c|l}
\hline
\textbf{Application} & \textbf{Domain} & \textbf{Datasets} & \textbf{Key Tools/Envs} & \textbf{Major Orch.} & \textbf{Model} \\ \hline

RAG~\cite{RAG-Survey} & Retrieval QA & WQA~\cite{RAG-WQA}, MS-MARCO~\cite{RAG-MSMARCO} & VecDB, embed & Pipeline & Qwen2.5-7B \\ \hline

HuggingGPT~\cite{shen2023hugginggpt} & Multimodal & TaskBench~\cite{shen2024taskbench} & Specialized models & Plan-Exec & DS-V4-Pro \\ \hline

\multirow{2}{*}{OpenDR~\cite{langchain-open_deep_research}} & \multirow{2}{*}{AI Search} & \multirow{2}{*}{YDC~\cite{DRS-YDC}, GAIA~\cite{DRS-GAIA}, HLE~\cite{DRS-HLE}} & \multirow{2}{*}{Search, VecDB, embed} & Parallel, Plan-Exec, & Qwen3.7-Max; \\
 & & & & Branching, Loop &  DS-V4-Pro, V4-Flash \\ \hline

Mini-SWE~\cite{mini-sweagent} & Coding & SWEBench Verified~\cite{jimenez2023swe} & FS, shell & ReAct & DS-V4-Pro \\ \hline

Codex~\cite{openai-codex} & Terminal & Terminal Bench~\cite{terminalbench} & FS, shell & ReAct & DS-V4-Pro \\ \hline

WebAgent~\cite{browsergym} & Browser & WebArena Verified~\cite{webarena-verified} & Browser & ReAct & Kimi-K2.6 \\ \hline

GUIAgent~\cite{osworld} & GUI & OSWorld~\cite{osworld} & Desktop & ReAct & Kimi-K2.6 \\ \hline

Claude Code~\cite{claudecode} & Tool-rich assistant & MCP-Atlas~\cite{mcpatlas} & MCP, search, FS, shell & ReAct & DS-V4-Pro \\ \hline

Openclaw~\cite{openclaw} & Office & WildClawBench~\cite{wildclaw} & MCP, search, FS, shell & ReAct & DS-V4-Pro \\ \hline

Pi-AutoR~\cite{pi-autoresearch} & AutoML/research & MLEBench~\cite{mlebench} & MCP, search, FS, shell & ReAct & DS-V4-Pro \\ \hline
\end{tabular}
\vspace{-3mm}
}
\end{table*}

\subsection{Application Collection}

\ABench addresses workload representativeness (R1) by selecting applications that exercise distinct systems-relevant regions of $W=(R,T,M,O)$.
Table~\ref{tab:suite-app-ds} summarizes the suite.
The goal is not to statistically sample all agents, but to preserve realistic task semantics while covering properties that shape latency, resource demand, state footprint, and cost.
The common workload model lets later observations be attributed to workload structure rather than ad hoc implementations.

For request distribution $R$, \ABench uses deliberately different task regimes rather than a flat domain list.
Short document-QA requests exercise small payloads and frequent queries, browser and GUI tasks introduce minute-scale interaction, and AutoML/research jobs stretch execution over longer horizons with larger artifacts.
Together, these cases cover scenario, payload-size, and duration variance.

For tools and environments $T$, the suite selects applications whose non-LLM components stress different resources and failure modes.
Retrieval and AI-search workloads exercise embedding, reranking, vector databases, and external search; coding and terminal workloads stress shell and filesystem execution; browser, GUI, and MCP-based workloads add stateful user interfaces, remote services, and network-facing I/O.
These examples cover tool-driven I/O behavior, state footprint, non-LLM bottlenecks, and cost profiles.

For orchestration $O$ and model choices $M$, \ABench covers both control-flow and inference-policy variance.
RAG represents a largely predetermined pipeline, interactive agents expose ReAct loops, and DeepResearch combines parallel execution, planning, branching, and loop-based refinement.
The suite also spans compact text models, stronger coding/tool-use models, multimodal models for browser and GUI interaction, and tiered DeepResearch inference.
These choices expose differences in execution length, state evolution, latency, context pressure, token behavior, and monetary cost.

\subsection{Serving Stack}

\ABench exposes serving-system factors (R2) through a modular serving stack with explicit boundaries between arrivals, orchestration, tools, models, and deployment.
At runtime, a generated request flows from the workload generator to an application orchestrator, which issues LLM calls, tool invocations, and environment interactions.
\ABench instruments this path at the orchestrator, proxy, sandbox, and container layers, linking each operation to its resource footprint.
The workload generator owns $R$: it samples tasks from the original datasets, applies an arrival process such as one-by-one, Poisson, or trace-driven arrivals, and submits requests through the orchestrator's protocol.
Each orchestrator owns $O$ and invokes configurable tools and environments ($T$) and model services ($M$).
This separation lets experiments vary arrivals, models, tools, or orchestration independently when studying factor sensitivity.
The guiding invariant is factor isolation: changing one factor should preserve the others whenever possible; a new arrival process reuses the same orchestrator, a new model endpoint preserves the same tool path, and a new tool deployment leaves the request distribution intact.
This invariant lets \ABench compare serving choices without confounding them with benchmark-specific glue code.

The deployment layer exposes $S{=}\langle H,C,A\rangle$.
\ABench packages runtime components as Docker containers so experiments can reproduce hardware allocation ($H$), component-serving mechanisms ($C$), and deployment architecture ($A$) under controlled resource limits and placement choices.
For example, a study can replace an SGLang instance with a vLLM instance to isolate the effect of the LLM-serving engine, or move a component from an embedded function call to an isolated online service to study disaggregation.
\ABench provides a lightweight programming framework for this conversion, requiring only code-level annotations and network configuration to switch between local invocation and remote procedure calls.

\subsection{Measurement Toolkit}

\ABench provides unified, component-level instrumentation (R3) that turns heterogeneous executions into comparable measurements of $Y=\Phi(W,S)$.
It records end-to-end latency, per-operation latency, input and output size, token usage, and monetary cost when available, along with component resource usage including CPU, memory, disk, network, and GPU usage.
These metrics expose where time, resources, tokens, and cost are spent under a given workload and serving configuration.

All instrumentation paths produce a normalized per-operation execution trace.
Each record captures the operation type, component, start and end time, input/output size, token usage, and monetary cost when available.
White-box applications whose source code can be modified (RAG, HuggingGPT, DeepResearch, Mini-SWE, WebAgent) emit these records through lightweight annotations.
Black-box third-party applications (Codex, GUIAgent, Claude Code, Openclaw, Pi-AutoR) are traced through LLM/tool proxies and sandbox hooks; for sandbox-heavy applications, \ABench uses SSH-based remote sandboxes to intercept actions through their hook methods.
Both paths feed the same trace schema, which matters because agents expose different internal interfaces: some expose Python call sites, while others expose only API traffic, shell actions, or sandbox hooks.
By treating these as collection paths for the same logical events, \ABench lets downstream analyses compare LLM calls, tool invocations, and environment actions across applications.

Containerized deployment lets \ABench attach production observability tools to each runtime component.
cAdvisor~\cite{cadvisor} collects CPU, memory, disk, and network usage, NVIDIA DCGM Exporter~\cite{dcgm} collects GPU utilization and memory usage, and Prometheus~\cite{prometheus} stores these metrics as time series.
After an experiment completes, \ABench's analysis library normalizes traces, extracts metrics, and generates either dedicated figures or customized Grafana~\cite{Grafana} dashboards.

\begin{figure}[t]
    \centering

    \includegraphics[width=1.0\linewidth]{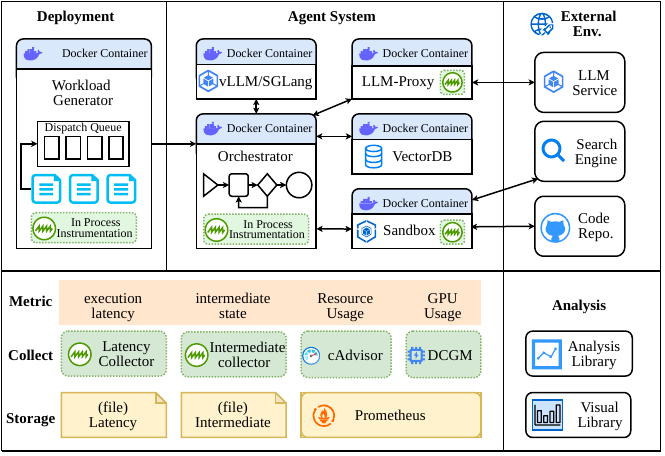}
    \caption{Suite Overview.}
    \label{fig:suite-overview}
    \vspace{-4mm}
\end{figure}

\subsection{Extensibility}

New orchestration frameworks, tool types, model capabilities, and deployment patterns emerge rapidly; to keep the suite current, \ABench supports extensibility by requiring new applications to describe themselves in the same workload and serving-system space as the existing suite.
To add an application, a user specifies its request distribution, tools and environments, models, orchestration, hardware resources, component-serving mechanisms, and deployment architecture.
Because the measurement toolkit already normalizes operation traces and resource time series, new applications do not require new metric definitions or attribution logic.

Integration is localized to the parts that actually differ.
A new workload can reuse existing arrival patterns, connect its orchestrator to the tool and model interfaces, and choose annotation-based or proxy-based tracing depending on whether its source code can be modified.
The deployment framework then gives the new application the same local-to-remote conversion used by the built-in applications.

\subsection{Default Benchmarking Settings}

Unless otherwise stated, the controlled characterization in Sections~\S\ref{sec:heavy}--\S\ref{sec:case-study} uses the following default deployment.
Each workflow except RAG runs on one cutting-edge NVIDIA GPU server, with each module placed in one Docker container and containers communicating through a shared-memory-based virtual network.
Mini-SWEAgent uses a self-deployed DeepSeek-V4-Pro served by SGLang v0.5.12, while the other non-RAG workflows use the Alibaba-Bailian API.
RAG runs on one x86\_64 server with 8$\times$ 4090D GPUs.

For tool deployment, embedding tasks use \textit{jina-embeddings-v3}, served by TEI~\cite{huggingface2025tei}.
RAG and DeepResearch deploy \textit{Milvus}~\cite{2021milvus} from its official Docker image.
DeepResearch uses the Exa search service.\footnote{\url{https://ai-sdk.dev/resources/tools/exa}.}
HuggingGPT serves machine learning models via HuggingFace Pipeline~\cite{huggingface2025pipeline}.

\section{Heavyweight with Non-LLM Dominance} \label{sec:heavy}

Agentic workloads differ fundamentally from conventional LLM serving along five systems dimensions:
end-to-end latency, token consumption, live state footprint, data movement, and monetary cost.
Using \ABench, we show that these workloads are heavyweight and stateful, and that
non-LLM components---sandboxes, retrieval engines, and environment interactions---frequently dominate both latency and cost.
Optimizing model inference alone therefore leaves the majority of execution time and operational expense untouched.

\subsection{Long Running and Non-LLM Bottleneck} \label{sec:long-execution}

\begin{figure}[tb]
    \centering
    \includegraphics[width=1.0\linewidth]{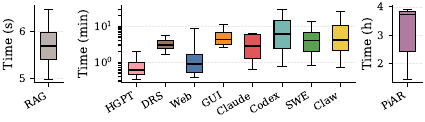}
    \caption{Distribution of end-to-end latency across requests for all ten \ABench applications.}
    \label{fig:e2e_delay_box}
    \vspace{-4mm}
\end{figure}

Conventional LLM serving workloads are sub-second, single-turn inference requests.
However, agentic workloads involve iterative, multi-step execution loops that interact with external environments.
To understand the latency characteristics of these workloads, we first measure the end-to-end execution latency
and analyze where the time is spent across different components.

Figure~\ref{fig:e2e_delay_box} shows the end-to-end latency distribution across all ten applications under sequential execution.
As shown by the figure, \textit{agentic executions are long-running and heavy-tailed.}
Unlike chatbot or microservice workloads that complete in sub-second to second timescales,
agentic requests span seconds to several hours. For example, coding tasks in \texttt{Mini-SWE} frequently
exceed ten minutes, and research tasks in \texttt{PiAutoResearch} can reach several hours.
\begin{figure}[tb]
    \centering
    \includegraphics[width=1.0\linewidth]{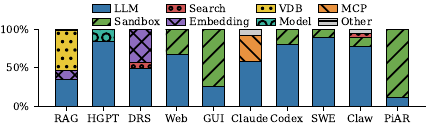}
    \caption{Decomposition of end-to-end execution time by component type.}
    \label{fig:execution_time_breakdown}
    \vspace{-3mm}
\end{figure}

To pinpoint the drivers of this extreme latency, we decompose the end-to-end execution time into component-level runtimes.
Figure~\ref{fig:execution_time_breakdown} decomposes execution time into LLM inference, sandbox operation, search,
databases operations, MCP calls, embedding, general model inference, and orchestration overhead.
As shown in the figure, \textit{non-LLM components dominate the critical path in half of the characterized applications.}
For example, the containerized desktop sandbox in \texttt{GUIAgent} accounts for over 70\% of total execution time,
while experimental runtimes consume 90\% in \texttt{Pi-AutoR}.

\noindent \textbf{Implications:}
These prolonged execution times fundamentally change system reliability and optimization requirements.
First, the extended runtime dramatically increases the likelihood of node or network failures during a single request,
making simple retry-on-failure strategies prohibitively expensive.
Serving systems must therefore provide \textit{lightweight, low-overhead fault tolerance} to checkpoint and resume agent sessions.
Second, because non-LLM components dominate the execution time, model-only optimizations yield diminishing returns.
The serving stack must transition to joint optimization to improve end-to-end latency.

\takeawaybox{Agent execution is long-running and non-LLM-dominated. Serving systems must manage tool latency and provide lightweight fault tolerance, not just optimize inference.}

\subsection{Increasing Token Usage} \label{sec:heavy-token}

\begin{figure}[t]
\centering
\includegraphics[width=1.0\linewidth]{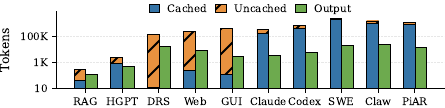}
\caption{Token consumption breakdown across characterized applications.}
\label{fig:token_breakdown}
\vspace{-3mm}
\end{figure}

Token consumption remains a primary operational cost driver for agentic applications. However, the structure of token usage in agent workloads differs fundamentally from single-turn LLM serving, characterized by super-linear context growth and unique prefix sharing. Figure~\ref{fig:token_breakdown} breaks down the per-session token consumption across our characterized applications.

\noindent \textbf{Multi-turn token amplification.}
Applications employing iterative reasoning loops (e.g., the ReAct pattern in \texttt{Mini-SWE} and \texttt{Codex}) accumulate tokens super-linearly across execution steps. Because each consecutive LLM call appends the entire historical conversation, tool execution logs, and system prompts to the context window, the input size scales quadratically with the turn count. Consequently, ReAct-style agentic workloads consume orders of magnitude more tokens per session than typical RAG workloads, which operate on fixed, single-turn prompts.

\noindent \textbf{Exploiting prefix cache dynamics.}
As context windows expand and multi-turn loops dominate, LLM inference engines must heavily rely on prefix caching to bypass redundant prefill computations. Our characterization shows that prefix cache hit rates vary dramatically across workloads, reflecting differences in prompt engineering and application workflows. Applications with static system prompts and append-only execution histories (e.g., \texttt{Claude Code}) achieve cache hit rates of up to 99\%. In contrast, applications that frequently restructure, reorder, or dynamically fetch context at each turn (e.g., \texttt{DeepResearch}) see prefix reuse drop to 1\% or lower.

\noindent \textbf{Implications:}
The dominance of cache-hit tokens makes prefix caching an absolute necessity rather than an optional optimization. To support these workloads, serving systems must treat context caches as first-class, long-lived resources. This directly shifts the systems challenge to managing the massive memory footprint of these cached performance states—a challenge we dissect in detail in \S\ref{sec:heavy-state}.

\takeawaybox{Token usage grows chronologically with agent complexity, heavily driven by Cache Hit tokens from iterative loops. Serving systems must optimize for high-volume context caching and token cost mitigation.}

\subsection{Substantial States to Manage} \label{sec:heavy-state}

\begin{figure}[t]
\centering
\includegraphics[width=1.0\linewidth]{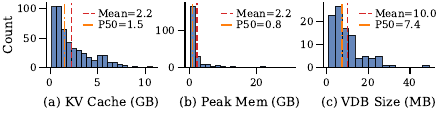}
\caption{Distribution of KV-cache performance state (a), active sandbox working-set memory (b), and persistent vector-database state (c) across representative sessions.}
\label{fig:state_size}
\vspace{-3mm}
\end{figure}

Unlike traditional stateless inference models, agentic execution models are highly stateful,
maintaining large execution contexts across their lifetime.
We distinguish cached performance state and persistent correctness state from
active working-set memory. The first two must be managed across session pauses;
the last is a transient resource demand and is not necessarily checkpointable
state at a quiescent boundary.

\noindent \textbf{Performance state.}
Performance state consists of cached intermediate artifacts that can be discarded
without violating execution correctness, but whose loss incurs significant recomputation latency.
The primary example is the LLM KV cache, which must be retained in GPU HBM or host DRAM
to avoid re-evaluating long, multi-turn contexts on subsequent steps.
As shown in Figure~\ref{fig:state_size}(a), as context lengths accumulate over dozens of iterations,
a single \texttt{Claude Code} session running a large model like DeepSeek-V4 can
consume up to 11\,GB of GPU memory for KV cache storage.
While evicting this state is semantically safe, because the inference engine can
reconstruct the cache by re-prefilling the history on the next call,
the steep latency and compute cost of prefill recomputation makes opportunistic eviction a costly trade-off.

\noindent \textbf{Persistent correctness state.}
Persistent correctness state comprises mutable execution data that
\emph{cannot} be lost without breaking the program's execution semantics.
In search-intensive applications like \texttt{DeepResearch}, each session instantiates an independent,
per-session vector database collection containing dynamically retrieved documents and embedding indices.
Figure~\ref{fig:state_size}(c) shows that the median collection size reaches 7.4\,MB, with larger sessions exceeding 49\,MB.
Coding agents such as \texttt{Codex} also require persistent filesystem changes,
installed dependencies, and process metadata. Agentic applications may further
make external side effects (e.g., sending an email), making the corresponding
external-component state part of correctness.

\noindent \textbf{Active working-set memory.}
A live sandbox also consumes DRAM while commands execute. As characterized in
Figure~\ref{fig:state_size}(b), median sandbox peak DRAM usage is about 0.8\,GB,
whereas the per-session peak reaches 28\,GB during compilation and unit tests.
This 28\,GB value is a working-set peak: transient compiler and test pages are
not necessarily persistent correctness state, nor do they all need to remain
live or be copied into a quiescent checkpoint.

\noindent \textbf{Implications:}
The co-existence of performance state, persistent correctness state, and a
large active working set complicates resource management.
Standard stateless microservices tolerate failures through simple, cheap request retries.
In contrast, failing or migrating an agentic session requires coordinating state checkpointing
across the LLM engine and tool runtimes, while resource provisioning must
separately accommodate transient sandbox peaks.

Crucially, the wide gap between typical and peak sandbox memory footprints (Figure~\ref{fig:state_size}(b))
exposes a key optimization opportunity: \emph{quiescent-point snapshotting}.
Initiating a sandbox checkpoint during periods of peak compilation activity is highly inefficient.
Instead, because agentic workflows naturally alternate between active execution steps and idle periods
(further analyzed in \S\ref{sec:idle-but-live}), scheduling sandbox checkpoints during inter-step idle intervals
(e.g., when the agent is waiting for LLM generation results) avoids treating a
transient working-set peak as the checkpointable live footprint.
Symmetrically, the host system can offload or page out the GPU KV cache of an idle session while the agent
is waiting for long-running sandboxed tool executions, maximizing active GPU memory utilization.

\takeawaybox{Agent sessions combine evictable KV caches, persistent correctness state,
and sandbox working sets that peak at 28\,GB.
State management should checkpoint persistent state at quiescent boundaries
and provision transient working-set peaks separately.}

\subsection{Extensive Data I/O} \label{sec:heavy-io}

\begin{figure}[t]
\centering
\includegraphics[width=1.0\linewidth]{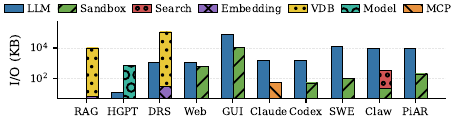}
\caption{Average and total data volume serialized, transmitted, and received per component type.}
\label{fig:all_agent_iosize}
\vspace{-3mm}
\end{figure}

The interactive nature of agentic workflows generates substantial volumes of intermediate data
that must be serialized and transferred across distributed components.
Figure~\ref{fig:all_agent_iosize} quantifies the average data volume received (left bars) and
sent (right bars) per component type, alongside the aggregate I/O per execution session.

Our analysis shows that a single agentic execution can generate and transfer tens of megabytes of data.
This extensive I/O is primarily driven by three sources.
First, \emph{sandboxed environments} incur substantial visualization and file synchronization overhead.
In \texttt{GUIAgent}, the desktop sandbox must capture and transmit high-resolution GUI screen frames
at each interaction step, generating the largest per-step data volume in our benchmark.
Second, \emph{embedding models and vector databases} produce substantial volumes of dense retrieval representations.
In our \texttt{DeepResearch} configuration, the embedding service emits
1,024-dimensional FP32 vectors. A request transfers over 20\,MB of serialized
embedding payload, corresponding to roughly 5,000 vectors (about five million
scalar embedding elements) before protocol overhead.
Third, \emph{LLM context accumulation} aggregates extensive state. While individual LLM text outputs are relatively compact,
multi-turn interactions require sending long system instructions, tools' responses, and historical observations.
In applications like \texttt{GUIAgent}, this repeated round-trip transfer accumulates to over 10\,MB of text data per session.

\noindent \textbf{Implications:}
This I/O is not merely a side effect of statefulness; it is a first-class bottleneck. In distributed deployments, serialization, deserialization, and network transmission of embeddings, sandbox observations, and accumulated context consume significant bandwidth and CPU cycles. Inter-component network capacity and serialization efficiency therefore constrain achievable concurrency and end-to-end latency, independent of compute provisioning.

\takeawaybox{Data movement reaches tens of MB per session and is driven by sandbox observations, embedding serialization, and context accumulation. This makes network bandwidth and serialization efficiency emerge as critical, non-negligible bottlenecks for high-concurrency agent-serving architectures.}

\subsection{Estimated Pay-as-you-go Cost Breakdown} \label{sec:heavy-cost}

\begin{figure}[t]
\centering
\includegraphics[width=1.0\linewidth]{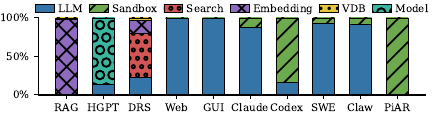}
\caption{Estimated per-request pay-as-you-go cost breakdown under public cloud pricing.}
\label{fig:monetary}
\vspace{-3mm}
\end{figure}

The physical characteristics detailed in the previous subsections---long execution runtimes, massive states,
and high I/O volumes---ultimately translate into financial costs.
While the system community heavily prioritizes model-centric metrics such as cost-per-million-tokens,
we show that the extensive non-LLM infrastructure required by agents fundamentally alters the economics of serving.
Figure~\ref{fig:monetary} estimates per-request pay-as-you-go operating cost under
public cloud pricing for LLM APIs (cache-read, prefill, decode), sandbox execution (E2B~\cite{e2b}),
and web search services (Firecrawl~\cite{Firecrawl_GitHub}).

We find that \textit{non-LLM infrastructure charges dominate this estimate in tool-heavy applications.}
In \texttt{Pi-AutoR}, sandbox charges constitute over 99\% of the per-request cost,
driven by GPU-accelerated containers for scientific simulations.

This operating-cost shift is fundamentally driven by resource idle times during sequential execution.
When dedicated sandboxes or retrieval environments are allocated to a session,
they continue to incur billing charges even when the session is waiting for long GPU-bound model inference steps,
leading to low resource utilization and high financial waste.

\noindent \textbf{Implications:}
Cost-aware serving systems must look beyond token efficiency and orchestrate the full tool-use lifecycle.
To operate agentic workloads cost-effectively at scale, serving platforms must deploy
active sandbox multiplexing and dynamic resource provisioning.
Similar to serverless cold-start optimizations, the serving stack should dynamically
pause, snapshot, and resume container environments during idle periods,
decoupling physical resource billing from the idle times of the agentic loop.

\takeawaybox{Non-LLM infrastructure dominates monetary cost in tool-heavy applications. Cost-aware serving must account for sandbox, retrieval, and environment expenses, not just token pricing.}

\section{Cross-Stack Heterogeneity} \label{sec:hetero}

Section~\ref{sec:heavy} established that non-LLM components dominate latency and cost,
implying that the serving stack must optimize the tool- and environment-heavy path rather than model inference alone.
Yet this path cannot be treated as a single target.
Agentic applications compose LLMs, small models, and diverse tools whose resource demands and runtime characteristics
differ fundamentally from one another.
Without explicit management, this heterogeneity causes resource fragmentation and performance interference.
We therefore study how heterogeneity manifests at three levels: components, tasks, and individual invocations;
and how it impacts the resource provisioning and scheduling of the serving system.

\subsection{Component-Level Heterogeneity} \label{sec:hetero-component}

Heterogeneity is most visible across components.
A single application spans GPU-bound LLMs, CPU-bound sandboxes (WebArena's browser, GUIAgent's desktop),
memory-bound vector databases, and network-bound search services (DeepResearch),
and it persists even within the GPU tier: HuggingGPT alone coordinates models spanning a
16.8-to-6032-GFLOP range~\cite{nsfw,podell2023sdxl}, each with a distinct memory footprint and latency profile.
The components also differ in lifecycle — stateful LLMs and long-running sandboxes demand sustained memory
and fast local storage, whereas stateless embedding models are ephemeral and elastically scalable.
On homogeneous hardware these divergent profiles fragment resources, saturating GPUs while CPUs sit idle
and precluding any static provisioning plan.
This component-level diversity is intuitive; the less obvious and more consequential heterogeneity emerges
within a single component — across tasks (\S\ref{sec:hetero-task}) and even across invocations of one task
(\S\ref{sec:hetero-invocation}) — which we examine next.

\subsection{Task-Level Heterogeneity} \label{sec:hetero-task}

\begin{figure}[tb]
  \centering
  \includegraphics[width=1.0\linewidth]{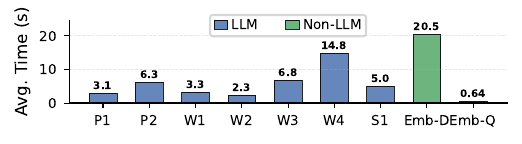}
  \caption{Execution time of tasks in the DeepResearch.}
  \label{fig:drs_time}
  \vspace{-3mm}
\end{figure}

Beyond component-level differences, tasks within the same application differ by up to 32$\times$
in latency even when served by the same component.
In DeepResearch, two LLM stages---\emph{write-search-plan}~(W2) and \emph{write-section}~(W4)---
issue to the same model, yet W4 is 6.43$\times$ slower (Figure~\ref{fig:drs_time}).
The gap is starker for embeddings: Embed-Doc is 32$\times$ slower than Embed-Query on the same model.

\begin{figure}[tb]
  \centering
  \includegraphics[width=1.0\linewidth]{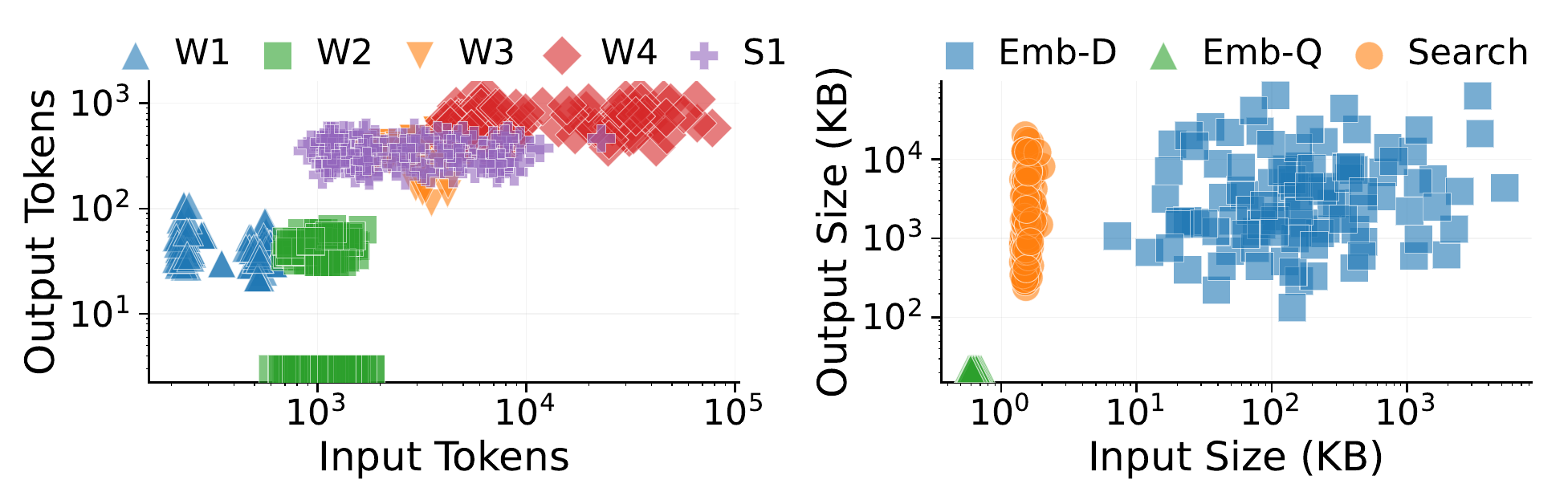}
  \caption{Workload of tasks in the DeepResearch.}
  \label{fig:drs_io}
  \vspace{-3mm}
\end{figure}

The driver is input/output scale, not the component itself.
For LLM tasks like W2 and W4, the execution cost is mainly determined by the number of
input tokens to prefill and the number of output tokens to decode.
As shown in Figure~\ref{fig:drs_io}(left), the average input tokens of W4 is 26$\times$ larger than W2's,
and the average output tokens of W4 is 16$\times$ larger than W2's, accounting for the 6.43$\times$ gap.
For embedding tasks, the execution cost is also mainly determined by the payload of texts to embed.
As shown in Figure~\ref{fig:drs_io}(right), Embed-Doc's payload is 325$\times$ larger than Embed-Query's.

\subsection{Invocation-Level Heterogeneity} \label{sec:hetero-invocation}

Figure~\ref{fig:drs_io} exposes a second layer of heterogeneity: the spread within each task is
itself large---per-call workload varies widely not only \emph{across} task types but also \emph{within} a single one.
This intra-task spread is not merely a consequence of aggregating different executions---even
within one agent execution, a task invoked repeatedly differs substantially from one invocation to the next.

\begin{figure}[tb]
    \centering
    \begin{tikzpicture}
      \node[anchor=south west, inner sep=0] (miniSweTraces) at (0,0) {
        \includegraphics[width=0.75\linewidth]{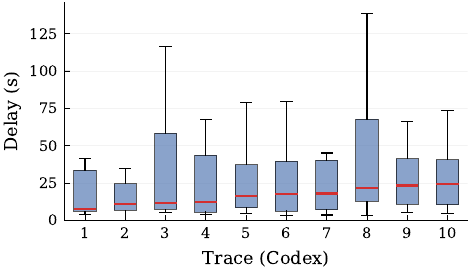}
      };
      \begin{scope}[x={(miniSweTraces.south east)},y={(miniSweTraces.north west)}]
        \node[anchor=south, fill=white, inner sep=0.5pt, font=\scriptsize]
          at (0.57,0.0) {Trace (Mini-SWE)};
      \end{scope}
    \end{tikzpicture}
    \caption{LLM-invocation latency across iterations for ten Mini-SWE traces.}
    \label{fig:sweagent_env_step_count_cdf}
    \vspace{-3mm}
\end{figure}

Figure~\ref{fig:sweagent_env_step_count_cdf} plots the latency distribution across iterations of ten Mini-SWE traces.
As shown in the figure, within a single trace the same LLM task varies by up to 30$\times$,
indicating remarkable intra-execution invocation-level heterogeneity.

\begin{figure}[tb]
  \centering
  \includegraphics[width=1.0\linewidth]{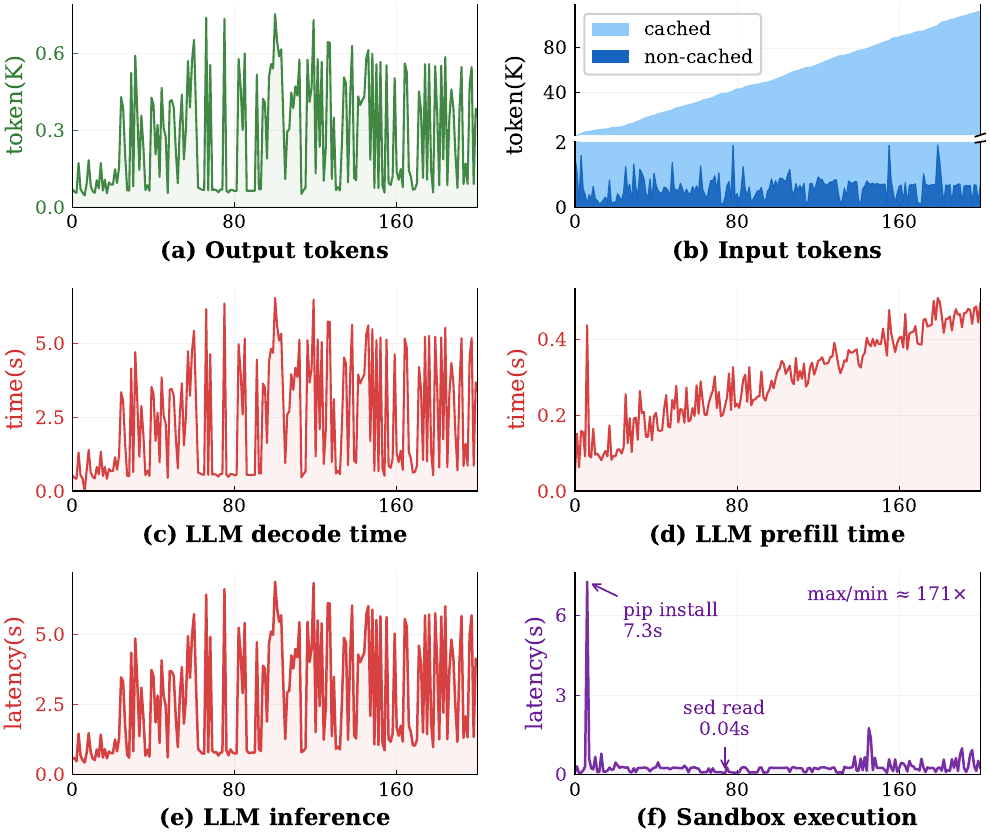}
  \Description{Per-iteration input (context), output, and latency across a single Mini-SWE-Agent execution.
  Input grows monotonically, yet latency fluctuates with output length rather than with the accumulated context.}
  \caption{Invocation-level heterogeneity in a single Mini-SWE execution: per-iteration input and output length, decomposed prefill and decode latency, and sandbox latency. Decode latency tracks output length, while prefill latency tracks the newly appended observation rather than the monotonically growing (cached) context.}
  \label{fig:sweagent_time_across_iterations}
  \vspace{-3mm}
\end{figure}

Similar to task-level heterogeneity, the invocation-level heterogeneity of LLM tasks
is also driven by per-invocation workload---specifically, the output it decodes and the input it must prefill.
We decompose the latency of LLM invocations into prefilling time and decoding time,
and plot the prefilling and decoding time of LLM invocations from a single Mini-SWE execution trace
in Figure~\ref{fig:sweagent_time_across_iterations}, together with the input length and output length of each invocation.

As shown in Figure~\ref{fig:sweagent_time_across_iterations}, the output length varies widely within a single trace,
contributing directly to the observed fluctuation in decoding latency, and thereby in end-to-end invocation latency.
Input length increases consistently because iterative agent loops append each action and observation to the conversation history.
Prefix caching reuses the KV states of this accumulated context and therefore avoids recomputing the cached tokens' projections
and feed-forward layers~\cite{zheng2024sglang}. Nevertheless, the cached prefix is not free: each newly appended token must
still attend over the preceding cached keys and values. If an invocation has $C$ cached tokens and $U$ uncached tokens,
the remaining attention work grows approximately with $U(C+U)$, and the attention kernel must read more cached KV state as
$C$ increases. The growing cached prefix therefore raises the global baseline of prefill latency across iterations.
Meanwhile, variation in $U$---the newly appended, non-cached observation---drives local fluctuations because these tokens
require full model computation and introduce additional attention queries. Thus, total context length explains the global
increase in prefill latency, while uncached input length better explains its iteration-to-iteration variation.

Besides LLM tasks, sandbox invocation also shows significant invocation-level heterogeneity.
We measured the latency of the sandbox invocation across iterations in the Mini-SWE execution trace.
As shown in Figure~\ref{fig:sweagent_time_across_iterations}, the sandbox invocation varies widely within a single trace,
with longest latency (\texttt{pip install} command) up to 171$\times$ slower than the shortest (\texttt{sed} command).

Invocation-level heterogeneity is not limited to iterative agent-loop applications. Other agentic applications
also exhibit significant invocation-level heterogeneity due to repeated calls to the same task within a single execution.
For example, in DeepResearch where each agent execution will trigger multiple search queries,
the search latency varies unpredictably across invocations: result payload sizes range from small snippets to multi-MB web pages,
and network conditions fluctuate when dozens of parallel queries compete for external bandwidth.
At the same time, the varying sizes of the web pages returned by search queries introduce varying payloads to the embedding task,
thereby contributing to varied embedding latency across invocations within the same execution.

\subsection{Performance Interference} \label{sec:hetero-interference}

The heterogeneity documented above has a direct operational consequence: heavyweight tasks can degrade
lightweight tasks that share the same serving resources. This interference arises through several related
mechanisms. First, a long-running request can cause \emph{head-of-line (HOL) blocking} when lightweight
requests cannot bypass it in a shared queue. Second, heterogeneous requests within the same batch can cause
\emph{co-batching interference}: the batch advances at the pace of the combined workload, so expensive requests increase
the iteration time observed by inexpensive ones. Third, concurrently executing tasks can contend for shared
physical resources, including CPU time, caches, memory bandwidth, and I/O. These mechanisms are neither
exhaustive nor mutually exclusive, but they capture three common ways in which cross-stack heterogeneity
translates into performance degradation. We isolate each mechanism with a controlled experiment.

\begin{figure}[tb]
  \centering
  \includegraphics[width=1.0\linewidth]{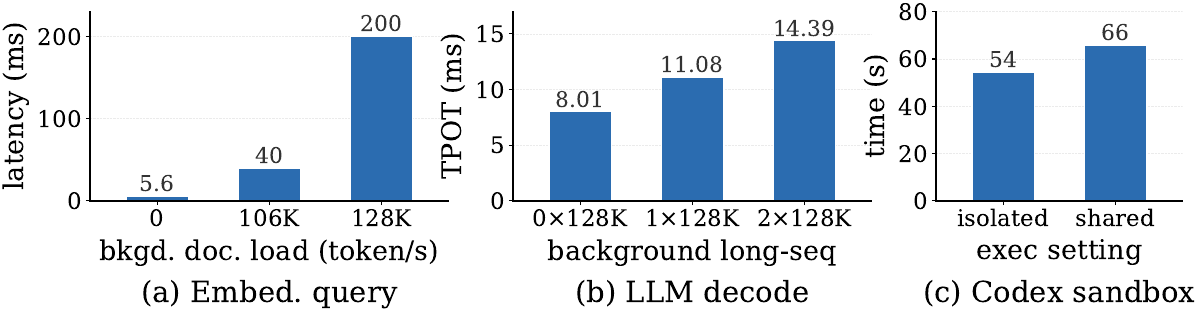}
  \Description{Mean embedding-query latency under increasing background document-embedding load; mean time per output token of a short-context LLM request when served alone or with one or two background 128K-context requests; and mean execution time of eight concurrent Codex sandboxes with isolated or shared CPU sets.}
  \caption{Three mechanisms of performance interference: HOL blocking in an embedding service (a), degradation of a short-context LLM request under increasing long-context background load (b), and shared-resource contention among sandbox operations (c).}
  \label{fig:interference}
  \vspace{-3mm}
\end{figure}

\noindent \textbf{Queueing interference.}
We first use the embedding service to demonstrate HOL blocking between tasks with different input sizes.
The lightweight \emph{embed-query} task and heavyweight \emph{embed-doc} task share one service instance and
one request queue. When served without background document traffic, an embed-query request completes in
5.6\,ms on average. As the background embed-doc load increases to 106K and 128K tokens/s, its average latency
rises to 40 and 200\,ms, respectively---slowdowns of 7.1$\times$ and 35.7$\times$. Large document requests
occupy the embedding worker for longer intervals, forcing latency-sensitive query requests to wait behind them.

\noindent \textbf{Co-batching interference.}
LLM engines batch active requests at each decoding iteration to improve GPU utilization~\cite{kwon2023efficient,zheng2024sglang}.
Each iteration computes one new token for every request in the batch, and its attention kernels read each
request's KV cache. Long-context requests therefore add substantially more KV-cache traffic and attention work
than short-context requests. Because requests in the batch advance through the same decoding iteration, this
additional work also increases the time per output token (TPOT) observed by short requests.
We fix one short-context (10K-token) request as the latency-sensitive victim and vary the number of co-served background
requests with 128K-token contexts. When served alone, the short request has a mean TPOT of 8.01\,ms.
Co-serving it with one long-context request raises TPOT to 11.08\,ms, a 38.3\% increase; with two long-context
requests, TPOT reaches 14.39\,ms, a 79.7\% increase (1.80$\times$). The monotonic slowdown shows that increasing
long-context background load delays even a fixed short-context request through shared decoding iterations.

\noindent \textbf{Shared-resource contention.}
Finally, we replay the \emph{install-klee-minimal} benchmark task in eight concurrent Codex sandboxes.
In the \emph{isolated} setting, each sandbox is pinned to a disjoint 16-CPU set. In the \emph{shared} setting,
all eight sandboxes are scheduled over the same pool of 128 CPUs. Although both settings use the same total
CPU capacity, sharing allows the sandboxes to interfere through CPU scheduling and the shared cache and memory
subsystems. Figure~\ref{fig:interference}(c) shows that mean execution time increases from 54 to 66\,s, a
1.22$\times$ slowdown. This experiment establishes shared-resource contention; attributing the slowdown to a
specific resource such as the last-level cache would additionally require hardware-counter measurements.

Together, these experiments show that heterogeneity creates interference at queueing, batching, and physical-resource
boundaries, motivating the task-aware serving explored in \S\ref{sec:case-study:task-dist}.

\takeawaybox{
Cross-stack heterogeneity causes HOL blocking, co-batching interference, and shared-resource contention when heterogeneous
tasks share serving resources. Serving systems must use task-aware queueing, batching, and isolation policies that
account for task size and resource demand.
}

\section{Shifting Bottlenecks} \label{sec:dynamic}

A natural follow-up question after establishing that agentic applications are
heavyweight and heterogeneous is: \emph{where} does the bottleneck lie?
In conventional LLM serving, the bottleneck is unambiguous: GPU inference dominates end-to-end latency.
In agentic serving, this is no longer the case.
As established in \S\ref{sec:characterization-bkg},
the measured system behavior $Y=\Phi(W,S)$ depends on both
the workload $W{=}\langle R,T,M,O\rangle$ and the serving system $S{=}\langle H,C,A\rangle$.
The bottleneck---the component dominating end-to-end latency---is one such observable $Y$
that shifts when changing any factor of $W$ or $S$.

This section presents empirical evidence for this phenomenon.
We first show that bottlenecks shift when workload factors change (\S\ref{sec:dynamic-workload}):
the request distribution $R$, the model choice $M$, the tool set $T$, and the orchestration structure $O$.
We then show that bottlenecks also shift when serving-system factors change (\S\ref{sec:dynamic-system}):
hardware resources $H$, component-serving mechanisms $C$, and deployment architecture $A$.
Together, these results demonstrate that the bottleneck location is not an intrinsic property of an application
but an emergent outcome of how that application is served under a given workload.

\subsection{Workload-Driven Shifting} \label{sec:dynamic-workload}

\noindent \textbf{Request distribution ($R$).}
Request distribution alone shifts the bottleneck through request type and payload,
even when all other factors are held constant.

\begin{figure}[tb]
    \centering
    \includegraphics[width=1.0\linewidth]{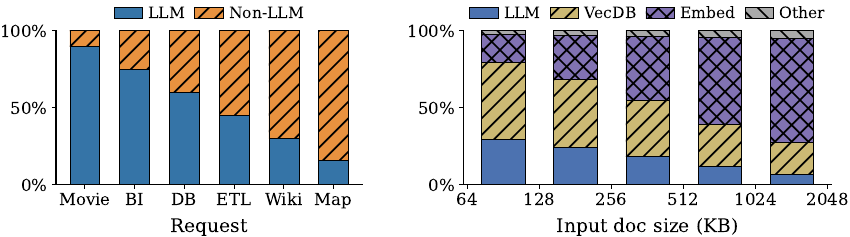}
    \caption{Effects of request type (left) and input-document size (right).}
    \label{fig:bottleneck_by_request}
    \vspace{-3mm}
\end{figure}

\emph{Request type.}
We run Claude Code on six request categories from MCP-Atlas~\cite{mcpatlas},
each solving a different problem with different MCP tools.
Figure~\ref{fig:bottleneck_by_request}(left) shows the latency breakdown.
The bottleneck differs sharply across request types:
LLM dominates in Movie, BI, and DB, reaching up to 90\% of total time,
whereas in ETL, Wiki, and MAP, tool execution dominates (up to 84\% of total time),
confirming that request type alone shifts the bottleneck.

\emph{Request payload size.}
Even within the same task type, payload-size variation shifts the bottleneck.
We isolate this effect with RAG, whose execution graph is fully predefined with no LLM-driven branching.
We drive it with MS-MARCO documents~\cite{RAG-MSMARCO} ranging from 64 to
2,048\,KB. Figure~\ref{fig:bottleneck_by_request}(right) reports the normalized
time breakdown for each document size. As documents grow, the
\techterm{embed\_doc} share increases while the \techterm{vdb\_store} and LLM
shares decrease, shifting the bottleneck from vector-database storage at small
inputs to embedding at large inputs.

\begin{figure}[tb]
    \centering
    \includegraphics[width=1.0\linewidth]{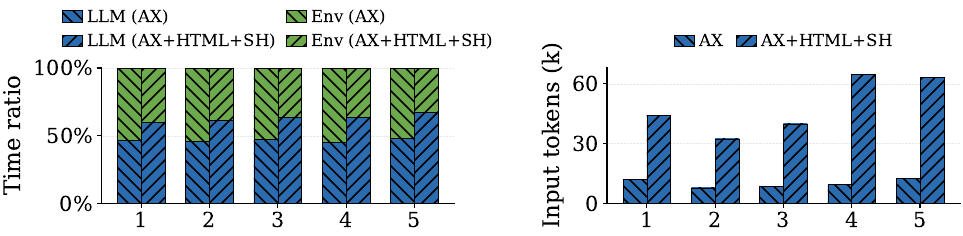}
    \Description{Latency breakdown of WebAgent under different observation-format configurations.
    Adding more observation formats increases LLM cost and shifts the bottleneck from browser interaction to LLM inference.}
    \caption{Effects of Tool set selection.}
    \label{fig:tool_bottleneck}
    \vspace{-3mm}
\end{figure}

\noindent \textbf{Tool set ($T$).} The tool set determines what observations the agent must process, and thus its LLM cost.
We demonstrate this with WebAgent, which observes browser state through three formats---accessibility tree, HTML, and screenshot---or combinations thereof: more formats reveal more state but lengthen the prompt sent to the LLM.
Running five web-browsing tasks under a single format (accessibility tree only) versus all three, we measure the per-task time breakdown (Figure~\ref{fig:tool_bottleneck}, left) and input tokens (right).
With a single format the observation is compact and browser interaction dominates; adding all three increases per-step input tokens by 4.8$\times$, raising LLM inference from 46.9\% to 61.6\% of total latency.

\begin{figure}[tb]
    \centering
    \includegraphics[width=1.0\linewidth]{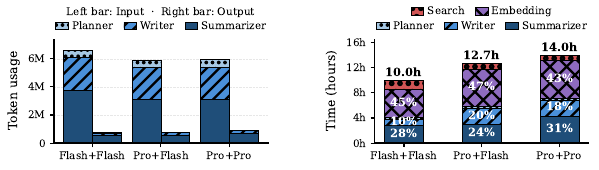}
    \Description{Input and output token usage by the planner, writer, and summarizer, together with absolute execution time and its component breakdown, for three DeepResearch writer--summarizer model configurations.}
    \caption{Token usage (left) and execution-time breakdown (right) of DeepResearch under three writer--summarizer configurations. The planner uses Qwen3.7-Max in all configurations.}
    \label{fig:model_selection_bottleneck}
    \vspace{-3mm}
\end{figure}

\noindent \textbf{Model selection ($M$).}
Model choice changes the per-token service time of LLM tasks and can therefore shift the application bottleneck
even when the execution graph and request set remain fixed. We demonstrate this with DeepResearch, which uses
separate LLM stages for planning, report writing, and summarization. The planner is fixed to Qwen3.7-Max, while
we vary the writer and summarizer between DeepSeek-V4-Flash and DeepSeek-V4-Pro. We evaluate three configurations,
with the writer listed first: \emph{Flash+Flash}, \emph{Pro+Flash}, and \emph{Pro+Pro}. In isolated
single-request measurements, V4-Flash generates 51 output tokens/s, whereas V4-Pro generates 21 output tokens/s,
making V4-Flash 2.43$\times$ faster in output-token throughput.

Figure~\ref{fig:model_selection_bottleneck} compares their token usage (left) and absolute execution-time breakdown
(right). With Flash+Flash and Pro+Flash, embedding remains the largest contributor, accounting for 45\% and 47\%
of total execution time, respectively; the aggregate LLM stages account for 38\% and 44\%. With Pro+Pro, the LLM
share rises to 49\%, overtaking embedding at 43\% and becoming the new bottleneck. Correspondingly, total execution
time increases from 10.0 hours with Flash+Flash to 12.7 hours with Pro+Flash and 14.0 hours with Pro+Pro.

The shift is driven primarily by this model-speed gap rather than token volume. Moving from Flash+Flash to Pro+Flash reduces
the total token volume, yet the LLM time share increases from 38\% to 44\%. Moving from Pro+Flash to Pro+Pro changes
token usage only modestly, while the LLM share increases further to 49\%. Thus, token counts alone do not predict the
latency breakdown: changing the serving speed of the writer and summarizer is sufficient to move the bottleneck from
embedding to LLM execution.

\begin{figure}[tb]
    \centering
        \includegraphics[width=1.0\linewidth]{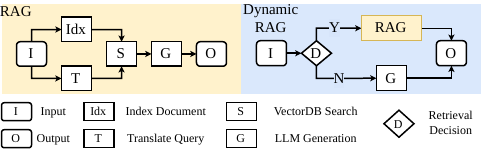}
        \caption{Orchestrator of RAG and RAG*.}
        \label{fig:rag-vs-ragstar-orch}
        \vspace{-3mm}
\end{figure}

\begin{figure}[tb]
    \centering
    \begin{minipage}[t]{0.45\linewidth}
        \centering
        \includegraphics[width=1.0\linewidth]{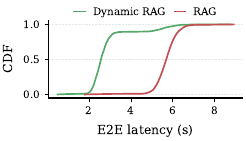}
        \caption{Orchestration effect comparison of RAG and RAG*.}
        \label{fig:rag-vs-ragstar}
        \vspace{-3mm}
    \end{minipage}
    \hfill
    \begin{minipage}[t]{0.45\linewidth}
        \centering
        \includegraphics[width=1.0\linewidth]{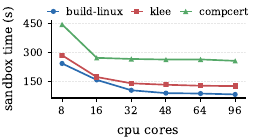}
        \caption{Codex sandbox with varied CPU allocation.}
        \label{fig:codex-varied-cpu}
        \vspace{-3mm}
    \end{minipage}
    \vspace{-3mm}
\end{figure}

\noindent \textbf{Orchestration ($O$).} Given the same task, same set of tools, and the same models,
different agent developers often design different orchestration structures.
The orchestration structure of an agentic application determines
how LLM and tools coordinate to process tasks, and thereby determines the execution graph.
We compare RAG (a static pipeline: embed$\to$retrieve$\to$generate) with Dynamic RAG (RAG*).
Figure~\ref{fig:rag-vs-ragstar-orch} shows the structural difference, where RAG* adds an
LLM-based retrieval-decision step that skips embedding and retrieval when deemed unnecessary.
Under the workload in Table~\ref{tab:suite-app-ds}, 90\% of queries bypass
retrieval in RAG*, leaving embedding idle and shifting the bottleneck from
embedding to LLM inference; Figure~\ref{fig:rag-vs-ragstar} reports the
resulting latency distributions.

\subsection{System-Driven Shifting} \label{sec:dynamic-system}

The bottleneck of an agentic application is not only determined by the workload factors ($R$, $T$, $M$, $O$),
but also influenced by the serving-system factors ($H$, $C$, $A$).
A change in serving configuration, such as varying GPU allocation, changing component-serving mechanisms,
or deploying in a co-located versus distributed manner, shifts the bottleneck.

\noindent \textbf{Hardware resources ($H$).}
The performance of a sandbox is strongly determined by its allocated resources,
especially when the sandbox is compiling or executing code.
We select ten TerminalBench tasks that involve expensive code execution under Codex Agent.
We vary each sandbox's allocation over 8, 16, 32, 48, 64, and 96 CPU
cores and measure all ten tasks. Figure~\ref{fig:codex-varied-cpu} shows three
representative tasks; allocating more cores decreases sandbox execution time
and can turn the bottleneck from sandbox execution to LLM inference.

\noindent \textbf{Component-serving mechanisms ($C$).} How a component is served
shapes its task latency and thus the application's latency breakdown.
For LLM serving alone, a developer chooses a framework (e.g., vLLM vs.\ SGLang),
an execution architecture (e.g., chunked-prefill vs.\ prefill--decode disaggregation),
and parameters such as batch size.
To illustrate, we serve GUIAgent's LLM with SGLang~\cite{zheng2024sglang}
and vary the batch size from 1 to 4.
With batch size 1, TPOT is 7~ms and tool execution (desktop interaction) dominates;
at batch size 4, decode contention raises TPOT to 30~ms,
making LLM inference 4$\times$ slower and turning it into the bottleneck---a shift
caused purely by an engine knob.

\noindent \textbf{Deployment architecture ($A$).} The deployment architecture governs
where components run and how they communicate: each may run on a server tuned to its needs,
co-locate with components of complementary resource demands for efficiency,
or co-locate with heavily communicating components to cut network overhead---
each choice yielding a different bottleneck.
We detail this in \S\ref{sec:case-study:co-location}, comparing co-located
and distributed deployments of RAG, with the LLM, embedding service,
and vector database placed on separate machines in the distributed case: under co-location,
network transfer is negligible and \emph{vector-db} dominates at 47\% of latency;
under distribution with request concurrency raised to 10, the network congests and transfer
grows to 67.5\% of total time.

\takeawaybox{
The bottleneck in agentic serving shifts whenever any workload factor ($R$, $T$, $M$, $O$)
or serving-system factor ($H$, $C$, $A$) changes.
Static profiling and fixed resource allocation cannot track these shifts;
serving systems must adopt online, per-request adaptation.
}

\section{Production-Trace Analysis}\label{sec:production}

The controlled experiments in Sections~\ref{sec:heavy}--\ref{sec:dynamic}
isolate workload and serving-system factors under laboratory conditions.
Production deployments add behavior that isolated benchmarks suppress:
user think time and approval, sessions that remain resumable across long
pauses, harness-generated context and auxiliary model calls, and repeated
tool requests across users.
We analyze 24-hour production traces from three deployed applications:
(1)~a \emph{coding agent} (35,037 sessions);
(2)~a \emph{search-based QA agent} (141,376 sessions); and
(3)~an \emph{Openclaw-like agent} (2,386 sessions).

These traces expose three properties that controlled benchmarks do not.
First, sessions spend far more time idle while retaining state than they
spend executing (\S\ref{sec:idle-but-live}).
Second, the agent control plane becomes a first-class cost center, consuming both
context capacity and model computation (\S\ref{sec:control-plane-tax}).
Third, repeated tool requests across sessions create a large, previously
hidden caching opportunity (\S\ref{sec:cross-request-redundancy}).

\subsection{Long Idle-but-Live Intervals}\label{sec:idle-but-live}

Benchmark executions advance to the next step as soon as the preceding
operation completes. Production sessions often pause while waiting for
human approval, a subagent result, or a follow-up instruction.
Such a session is logically live but may be locally inactive: its sandbox,
terminal context, KV cache, and conversation history must remain available
for resumption even while the session issues no new LLM or tool work.

\begin{figure}[tb]
    \centering
    \includegraphics[width=\linewidth]{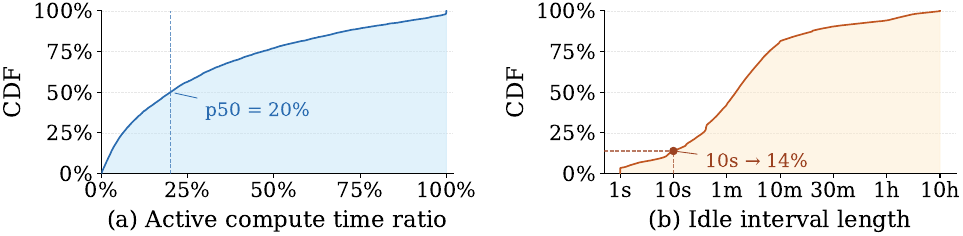}
    \Description{(a)~CDF of the ratio of active execution time to total
    session lifetime. Most sessions spend only a small fraction of their
    lifetime executing. (b)~Distribution of idle-interval lengths across
    production sessions.}
    \caption{Idle-but-live behavior in production coding-agent sessions.}
    \label{fig:session_idle_live}
    \vspace{-3mm}
\end{figure}

Figure~\ref{fig:session_idle_live}(a) plots, for the 35,037 sessions, the
ratio of time attributed to LLM and tool execution to total session
lifetime. The median session executes for only 20\% of its lifetime, and
70\% of sessions execute for less than half of their lifetime.
Figure~\ref{fig:session_idle_live}(b) shows that idle intervals range from
seconds---for example, while waiting for a tool result or human approval---to hours,
including overnight pauses. Most intervals fall between 1 and 10 minutes.
Throughout these intervals, the per-session sandbox that holds
correctness-critical state remains allocated.

\noindent \textbf{Implications:}
A binary ``running'' versus ``finished'' lifecycle is insufficient for
agentic serving. The harness should expose a third, quiescent
``waiting'' state and, when available, the expected resume trigger or
deadline. The serving layer can then manage correctness state and
performance state separately: checkpoint or offload sandbox and terminal
state to durable tiers, while retaining, migrating, or evicting KV cache
according to its expected reuse benefit. Resume prediction and prefetch
can hide restoration latency. We evaluate one such policy in
Section~\ref{sec:case-study:pro-scheduling}.

\takeawaybox{
Production agents are often waiting, not finished. Treating ``waiting''
as a first-class lifecycle state enables aggressive state reclamation
without sacrificing resumability.
}

\subsection{The Agent Control-Plane Tax}\label{sec:control-plane-tax}

A production agent runs inside a \emph{harness} that implements its application-level
control plane: it assembles prompts, manages the context window, drives and
guards the tool-calling loop, and preserves state across a long-running session.
This machinery is necessary for multi-step execution. We use \emph{agent
control-plane tax} to denote the incremental serving cost it introduces beyond
goal-directed model and tool execution. The tax has three forms: control-plane
state occupies context-window capacity, auxiliary tasks invoke additional LLM
work, and the interaction between session idleness (\S\ref{sec:idle-but-live})
and cache lifetime causes repeated prefill.

\begin{figure}[tb]
    \centering
    \includegraphics[width=\linewidth]{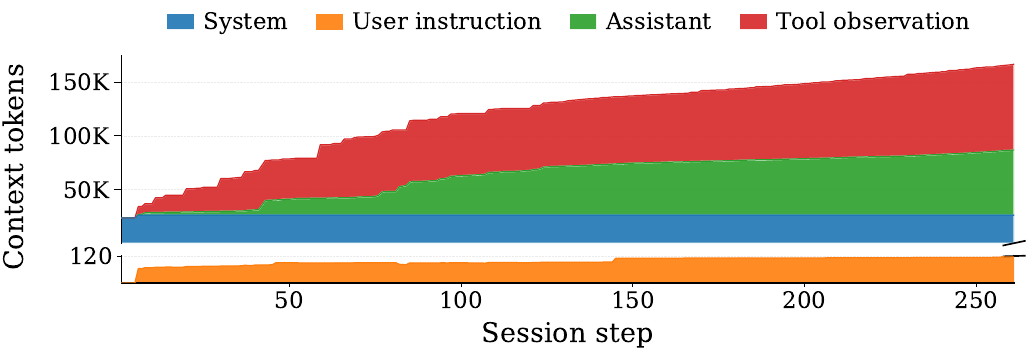}
    \Description{Stacked context composition across the LLM calls of one
    production session, separating system messages, user instructions, and
    accumulated history.}
    \caption{Composition of the LLM input context over one production session.}
    \label{fig:llm_tax}
    \vspace{-3mm}
\end{figure}

\noindent \textbf{Context-capacity tax.}
We partition the input assembled for each LLM call into three semantic
categories.
\emph{System messages} contain the largely static system prompt, including
the role definition, tool and skill descriptions, memories, and framework
directives such as date-time reminders and session settings.
\emph{User instructions} contain the user's requests, which are typically
short but may occur multiple times in one session.
\emph{History} contains the accumulating execution trajectory: the model's
reasoning, its actions (tool calls), and the resulting observations (tool
outputs).

Figure~\ref{fig:llm_tax} tracks this composition in an example production
session that contains multiple user interactions. At the first step,
system messages account for 99.7\% of the tokens in the assembled input.
As the session proceeds, history---dominated by tool calls and observations---
grows rapidly and eventually accounts for 84.3\% of the input.
At the final step (step-261), the model emits only 151 tokens for its next action
but must first process 166,721 context tokens.
Thus, even a short control decision can require a very large prefill.
Not every historical token is avoidable; rather, this breakdown quantifies the
context capacity and prefill work required to carry the control state forward.

Production agents control this growth through \emph{context compaction},
triggered automatically by the agent harness or manually by the user.
Compaction asks the model to summarize the current context into a shorter
replacement that should preserve the information needed by future steps.
We report the post-compaction context length divided by the pre-compaction
length; lower ratios indicate greater compression.

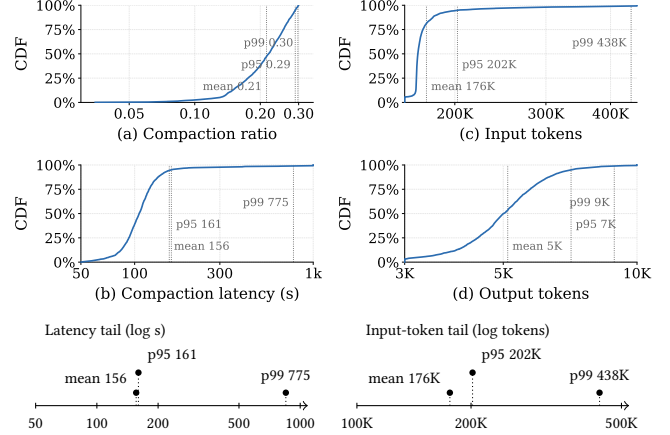
\begin{figure}[tb]
    \centering
    \includegraphics[width=\linewidth]{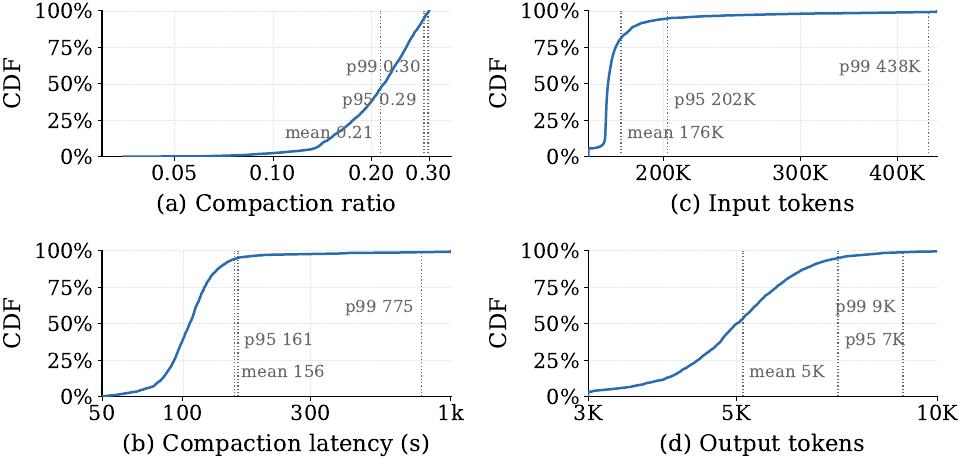}
    \par\vspace{1mm}
    \begin{tikzpicture}[x=1cm,y=0.38cm,font=\tiny]
      \begin{scope}
        \node[anchor=west] at (0,2.6) {Latency tail (log s)};
        \draw[->] (0,0) -- (3.6,0);
        \foreach \x/\lab in {0/50,0.81/100,1.62/200,2.69/500,3.5/1000} {
          \draw (\x,0.10) -- (\x,-0.10) node[below] {\lab};
        }
        \draw[densely dotted] (1.33,0) -- (1.33,0.45);
        \fill (1.33,0.45) circle (1.2pt) node[above left] {mean 156};
        \draw[densely dotted] (1.36,0) -- (1.36,1.15);
        \fill (1.36,1.15) circle (1.2pt) node[above right] {p95 161};
        \draw[densely dotted] (3.31,0) -- (3.31,0.45);
        \fill (3.31,0.45) circle (1.2pt) node[above] {p99 775};
      \end{scope}
      \begin{scope}[xshift=4.25cm]
        \node[anchor=west] at (0,2.6) {Input-token tail (log tokens)};
        \draw[->] (0,0) -- (3.6,0);
        \foreach \x/\lab in {0/100K,1.51/200K,3.5/500K} {
          \draw (\x,0.10) -- (\x,-0.10) node[below] {\lab};
        }
        \draw[densely dotted] (1.23,0) -- (1.23,0.45);
        \fill (1.23,0.45) circle (1.2pt) node[above left] {mean 176K};
        \draw[densely dotted] (1.53,0) -- (1.53,1.15);
        \fill (1.53,1.15) circle (1.2pt) node[above right] {p95 202K};
        \draw[densely dotted] (3.21,0) -- (3.21,0.45);
        \fill (3.21,0.45) circle (1.2pt) node[above] {p99 438K};
      \end{scope}
    \end{tikzpicture}
    \Description{(a)~Distribution of post-compaction to pre-compaction context ratios.
    (b--d)~Latency, input tokens, and output tokens per compaction event. Log-scale
    tail summaries expose the latency and input-token p99 values.}
    \caption{Context compaction: post-/pre-compaction context ratio (a), latency (b), input tokens (c), and output tokens (d). The lower log-scale tail summaries show the off-axis p99 values of 775\,s and 438K input tokens.}
    \label{fig:compaction}
    \vspace{-3mm}
\end{figure}

Across the 35,037 sessions, we observe 3,170 compaction events.
Figure~\ref{fig:compaction}(a) shows that 99\% of events reduce the
context by more than 70\%, indicating that a substantial fraction of
production context is compressible rather than requiring verbatim replay.

\noindent \textbf{Auxiliary-compute tax.}
Compaction itself is not free. As shown in
Figure~\ref{fig:compaction}(c,d), one event consumes, on average, 176K input
tokens and generates 5K output tokens. When the previous prefix remains
cached, most input tokens can avoid recomputation, but producing the
compacted summary still requires a long decode. As shown in
Figure~\ref{fig:compaction}(b), across the 3,170 events,
compaction latency averages 156\,s, with p95 and p99 latencies of 161\,s and
775\,s, respectively.

The harness also invokes LLMs for auxiliary tasks outside the main reasoning-and-action loop,
including safety guardrails that vet actions and loop
detection that identifies non-progressing executions. Beyond compaction,
these auxiliary tasks collectively add 2,684 LLM calls, 6.5M input tokens,
and 50K output tokens across the 35,037 sessions.

\begin{figure}[tb]
    \centering
    \includegraphics[width=\linewidth]{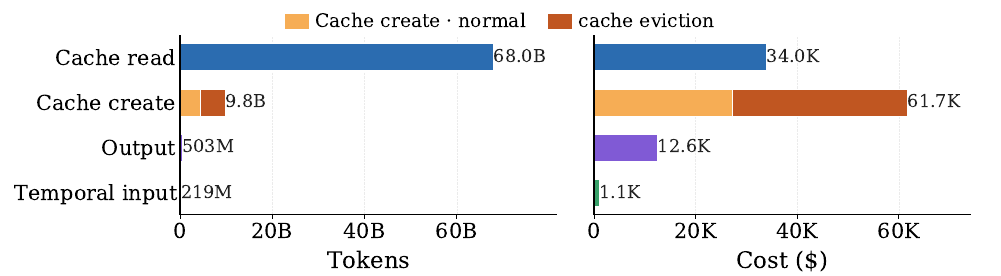}
    \Description{Token and monetary-cost breakdowns by cache-read, cache-create,
    output, and temporal-input categories, with cache-create cost attributed to normal operation and cache eviction.}
    \caption{Token and monetary-cost impact of prefix-cache eviction.}
    \label{fig:cache_eviction}
    \vspace{-3mm}
\end{figure}

\noindent \textbf{Cache-reprefill tax.}
Although each LLM input is large, the harness normally grows it from the
previous input, preserving prefix continuity so that the serving engine
can reuse the KV cache. Retaining long-context KV state is expensive,
however, so production cache entries have a finite time-to-live (TTL)---5
minutes in our deployment. After the TTL expires, the next call must
re-prefill the evicted prefix. This fixed TTL is poorly matched to
human-paced agent sessions: the 1--10 minute idle intervals observed in
\S\ref{sec:idle-but-live} frequently overlap or exceed it.
Across the 35,037 sessions, 59.4\% experience at least one eviction event.
We break down their total token usage into the four categories reported by the deployment:
\emph{cache-read}, \emph{cache-create}(tokens whose KV-cache will be cached),
\emph{temporal-input}(tokens whose KV-cache will be discarded), and \emph{output}.
We estimate the corresponding monetary cost breakdown using the stated prices
for the Claude Opus 4.6 model (\$0.5/MToks, \$6.25/MToks, \$5/MToks, \$25/MToks, respectively).
As shown in Figure~\ref{fig:cache_eviction}, cache evictions contribute 55.9\% of the
total \emph{cache-create} tokens and ultimately account for 31.5\% of aggregate monetary cost across these sessions.

\noindent \textbf{Implications:}
The agent control-plane tax is a cross-layer systems problem, not merely a
prompt-engineering issue. First, tool interfaces should be
\emph{agent-native}. Rather than reinjecting full schemas and raw outputs
on every turn, tools should expose stable schema identifiers, typed and
size-bounded observations, incremental deltas, and retrievable handles to
full artifacts. The harness can then materialize only the fields needed
for the current decision, reducing token volume and prefill while
preserving exact state outside the context window. Second, context should
be virtualized as tiered state. Structured histories, hierarchical
summaries, and retrieve-on-demand detail can keep cold observations out
of the prompt and replace monolithic, near-limit compaction. Third, the
harness and model server should coordinate cache policy: fixed-TTL
eviction should give way to reuse-aware KV retention, offloading, and
prefetch based on the session's waiting state and predicted return.
Auxiliary calls should likewise be tagged by purpose so the serving
system can account for them separately and, where correctness and safety
permit, batch, cache, or route them to lower-cost models.

\takeawaybox{
The agent control plane is a first-class cost center: repeated schemas and
observations occupy context, auxiliary calls add model work, and idle gaps
cause cached prefixes to be re-prefilled. Reducing this tax requires
agent-native tool observations, semantic context virtualization, and
reuse-aware KV-cache management.
}

\subsection{Exploitable Cross-Request Redundancy}\label{sec:cross-request-redundancy}

Single-request benchmarks cannot reveal reuse across users and sessions.
We quantify such reuse at two external-tool boundaries.
The search-based QA agent issues a web search before answering each user
request. Its 141,376 sessions generate 373,678 search invocations, whose
mean, median, and p90 latencies are 1.349~s, 1.258~s, and 1.708~s,
respectively. Search accounts for a median of 35\% and a mean of 52\% of
session end-to-end latency.
The Openclaw-like office-automation agent sometimes fetches the full
content of a URL. We analyze the 2,386 sessions that issue at least one
fetch, comprising 4,389 fetch invocations with mean, median, and p90
latencies of 2.657~s, 1.482~s, and 4.987~s. Fetching accounts for a median
of 2.9\% and a mean of 6\% of session latency.

\begin{figure}[tb]
    \centering
    \includegraphics[width=\linewidth]{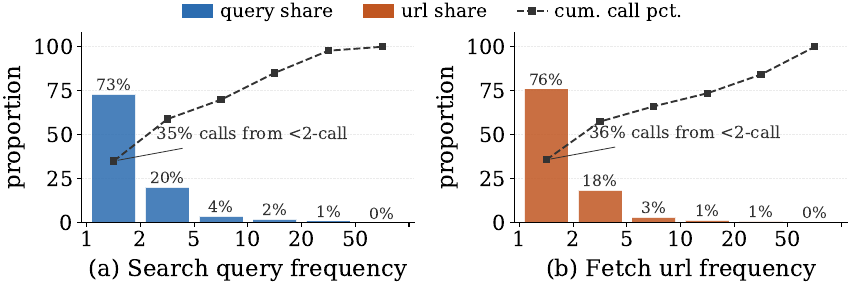}
    \Description{Distribution of within-day invocation frequencies for
    distinct search queries and fetched URLs, showing substantial reuse
    across sessions.}
    \caption{Cross-session redundancy in search queries and URL fetches.}
    \label{fig:cross_request_redundancy}
    \vspace{-3mm}
\end{figure}

For each distinct search query and fetched URL, we count its invocations
within the 24-hour trace. Figure~\ref{fig:cross_request_redundancy} shows
that 27\% of distinct search queries recur across sessions, and these
recurring queries account for 67.3\% of all search invocations. In the
Openclaw-like agent, 24\% of distinct URLs recur and account for 64\%
of all fetch invocations. Thus, a minority of distinct tool inputs is
responsible for a large fraction of external calls, repeatedly adding
latency and consuming network bandwidth.

\noindent \textbf{Implications:}
Cross-session reuse calls for caching at the shared tool-serving boundary,
where reuse can be aggregated across agents rather than hidden inside one
session. A first tier can cache exact query-to-result mappings; a second
can deduplicate URL fetches and reuse the fetched object (and, when
applicable, its parsed representation).
This mechanism requires no change to the agent's control logic.
We evaluate a two-tier design in Section~\ref{sec:case-study:caching}.

\takeawaybox{
A minority of distinct queries and URLs generates a large fraction of
production tool calls. Freshness- and tenancy-aware shared caches can
remove substantial latency and cost that per-session optimization cannot
see.
}

\section{Design Explorations}\label{sec:case-study}

The characterization in
Sections~\S\ref{sec:heavy}--\S\ref{sec:cross-request-redundancy} identifies
concrete bottlenecks and inefficiencies in agentic serving.
We now translate four of these findings into proof-of-concept design
explorations, each targeting a specific property:
cross-stack heterogeneity motivates \emph{task-aware serving}
(\S\ref{sec:case-study:task-dist}) and \emph{communication-aware placement}
(\S\ref{sec:case-study:co-location});
idle-but-live sessions motivate \emph{state offloading}
(\S\ref{sec:case-study:pro-scheduling});
and cross-request redundancy motivates \emph{tool-result caching}
(\S\ref{sec:case-study:caching}).
These are not components of a monolithic system; they are independent,
characterization-guided interventions that each demonstrate significant
gains over workload-oblivious baselines.

\begin{figure*}[tb]
    \centering
    \includegraphics[width=\textwidth]{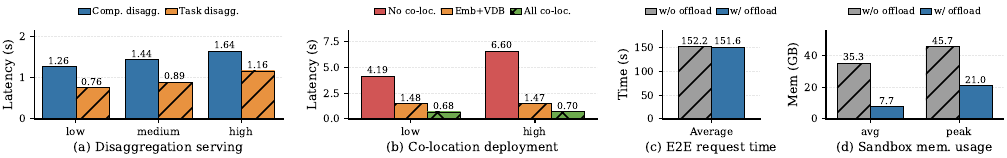}
    \caption{Effectiveness of task-aware serving, communication-aware placement, and state offloading}
    \label{fig:case-study}
\end{figure*}

\subsection{Task-Aware Serving} \label{sec:case-study:task-dist}

Section~\S\ref{sec:hetero} shows that tasks sharing the same component
frequently differ by orders of magnitude in input size, latency, and
resource footprint.
Co-serving these heterogeneous tasks in a single worker pool causes
head-of-line blocking and performance interference.
Task-disaggregated serving addresses this by deploying each logical task
(e.g., \techterm{embed-query}, \techterm{embed-doc}, \techterm{llm-judge})
as an independent service with dedicated resources, enabling performance
isolation, task-specific optimization, and fine-grained independent scaling.

We evaluate task-disaggregated deployment on Dynamic RAG under low, medium,
and high request rates (0.5$\times$, 0.7$\times$, 0.9$\times$ of peak
throughput).
As shown in Figure~\ref{fig:case-study}(a),
task disaggregated deployment reduces average latency by 40\%, 38\%, and
29\% under the three load levels, respectively, compared to a
component-sharing baseline with the same GPU count.
The improvement stems primarily from mitigating queueing delays and
interference between heterogeneous tasks sharing the same embedding and
LLM instances.

\subsection{Communication-Aware Placement} \label{sec:case-study:co-location}

Task disaggregation improves isolation but introduces a placement
challenge: naively assigning each task type to a dedicated machine
incurs prohibitive network overhead from massive intermediate states
(Section~\S\ref{sec:heavy}) and wastes resources due to heterogeneous
hardware demands (Section~\S\ref{sec:hetero-component}).
Communication-aware co-location places high-communication tasks
(e.g., embedding models and vector databases) on shared servers to
minimize data transfer, while pairing tasks with complementary resource
profiles (e.g., GPU-bound LLMs and memory-bound vector databases) to
improve hardware utilization.

We deploy RAG on a multi-node cluster and compare three placement
strategies: (1)~\techterm{co-none}, each component on a dedicated server;
(2)~\techterm{co-vdb-embed}, vector database and embedding tasks
co-located; (3)~\techterm{co-all}, all tasks on one server.
Figure~\ref{fig:case-study}(b) shows that co-locating vector
database and embedding tasks reduces average latency by 2.8$\times$ and
4.5$\times$ under low and high load, respectively.
Detailed profiling confirms that network communication accounts for
67.5\% of execution time in the \techterm{co-none} baseline but becomes
negligible with co-location.

\begin{figure}[tb]
    \centering
        \includegraphics[width=\linewidth]{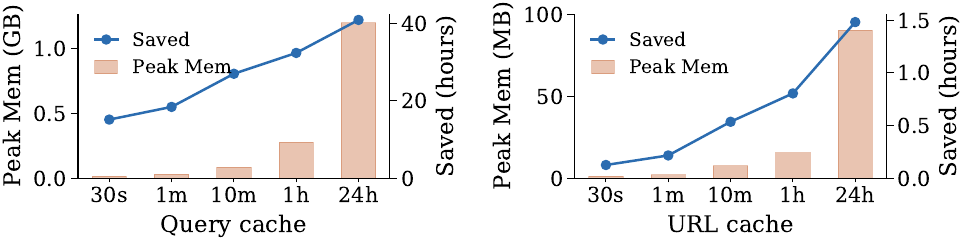}
        \Description{Latency reduction from tool-result caching.}
        \caption{Effectiveness of tool-result caching.}
        \label{fig:principle:caching}
    \vspace{-3mm}
\end{figure}

\subsection{State Offloading} \label{sec:case-study:pro-scheduling}

Section~\S\ref{sec:idle-but-live} shows that production sessions spend the
majority of their lifetime idle but holding expensive state---sandbox
environments, terminal contexts, KV cache, and conversation history.
Eagerly reclaiming this state risks breaking resumability; keeping it
allocated wastes memory proportional to the number of paused sessions.
Our controlled Mini-SWE experiment is a related proof of concept for
\emph{inter-component} idle intervals, rather than an evaluation driven by
the human-paced production pauses in \S\ref{sec:idle-but-live}. During each
LLM planning call, the sandbox is inactive and can be offloaded until the
next tool step.

Figure~\ref{fig:sweagent_time_across_iterations} shows a context-dependent
upward trend in the prefill baseline, but total LLM latency remains variable
because output length and newly appended input fluctuate across iterations.
We therefore do not assume monotonic latency or use context length as a
duration predictor. The prototype uses elapsed time of the in-flight planning
call as its trigger: once that time exceeds a fixed threshold, it offloads the
sandbox and restores it before returning control to the next tool step.

As shown in Figure~\ref{fig:case-study}(c, d), proactive offloading
reduces average and peak memory consumption by 4.6$\times$ and
2.1$\times$ respectively, with latency increasing within 0.5\%. The memory
bars report aggregate resident memory across all concurrently active sandbox
processes in this Mini-SWE run, not a per-session peak; they are therefore not
directly comparable to the 28\,GB per-session working-set peak in
Figure~\ref{fig:state_size}(b).

\subsection{Tool-Result Caching} \label{sec:case-study:caching}

Section~\S\ref{sec:cross-request-redundancy} reveals heavy cross-request
redundancy in production agentic workloads: in a search-based QA
application, recurring queries account for a large share of all search
API calls, while in an Openclaw-like agent, duplicate URL fetches waste
network bandwidth.
This redundancy creates a high-leverage caching opportunity that requires
no modification to agent logic or tool interfaces.

We evaluate a two-tier tool-result cache using the production traces from
Section~\S\ref{sec:cross-request-redundancy}.
The first tier performs exact query matching: when an incoming search
query matches a previously seen query, the cached result is returned
directly.
The second tier performs URL-level deduplication: when a web-fetch
request targets a URL whose content was already retrieved within a
configurable staleness window, the cached page content is reused.
Figure~\ref{fig:principle:caching} shows the results.
With a 10-minute TTL, the query cache eliminates 35.2\%
of redundant search calls in the search-based QA agent, saving 27 hours of
aggregate search latency (19.3\% of the total).
Similarly, with a 10-minute TTL, the URL cache eliminates 11.65\%
of redundant web-fetch calls in the Openclaw-like agent, saving 32 minutes
of aggregate fetch latency (16.5\% of the total).

\section{Conclusion}

We presented \ABench, a benchmark suite of ten agentic applications and a
modular serving stack for systems-level study. Controlled experiments and
production-trace analysis reveal that agentic workloads are heavyweight,
cross-stack heterogeneous, and dynamically shifting, while production
sessions further exhibit long idle-but-live intervals, an LLM
control-plane tax, and heavy cross-request redundancy---properties
invisible to model-centric benchmarks. Four characterization-guided
design explorations---task-disaggregated serving, communication-aware
placement, state offloading, and tool-result caching---yield
29--40\% lower latency, up to a 4.5$\times$ end-to-end speedup over
fully distributed placement,
4.6$\times$ less memory, and substantial savings in redundant external
calls, demonstrating that workload-aware coordination of models, tools,
and state is essential for efficient agent serving. \ABench will be
released as open source.

\bibliographystyle{ACM-Reference-Format}
\bibliography{reference.bib}

\end{document}